\documentclass[aps,pre,onecolumn,11pt,notitlepage,nofootinbib]{revtex4-1}

\usepackage{bm,epsfig,graphics,amssymb,amsmath,subeqnarray,setspace,graphicx,amsthm,epstopdf,subfigure,color,txfonts}
\usepackage{setspace,natbib}
\usepackage{multirow}
\usepackage{epsfig,graphics,amssymb,amsmath,color,setspace}
\usepackage{graphicx}
\usepackage{natbib}
\usepackage[sfdefault=cmbr]{isomath}

\newcommand{\reftext}{}
\newcommand{\tcb}{}

\begin{document}
\title{A generalized traction integral equation for Stokes flow,
with~applications to near-wall particle mobility and viscous~erosion}
\author{William H. Mitchell and Saverio E. Spagnolie}
\affiliation{Department of Mathematics, University of Wisconsin-Madison, 480 Lincoln Dr., Madison, WI 53706}%
\date{\today}

\begin{abstract}        
A double-layer integral equation for the surface tractions on a body
moving in a viscous fluid is derived \tcb{which allows} for the
incorporation of a background flow and/or the presence of a plane wall.
The Lorentz reciprocal theorem is used to link the surface tractions on
the body to integrals involving the background velocity and stress
fields on an imaginary bounding sphere (or hemisphere for wall-bounded
flows). The derivation requires the velocity and stress fields
associated with numerous fundamental singularity solutions which we
provide for free-space and wall-bounded domains. Two sample applications
of the method are discussed: we study the tractions on an ellipsoid
moving near a plane wall, which provides a more detailed understanding
of the well-studied glancing and reversing trajectories \tcb{in the
context of particle sedimentation}, and the erosion of bodies by a
viscous flow, in which the surface is ablated at a rate proportional to
the local viscous shear stress. Simulations and analytical estimates
suggest that a spherical body in a uniform flow first reduces nearly but
not exactly to the drag minimizing profile and then vanishes in finite
time. The shape dynamics of an eroding body in a shear flow and near a
wall are also investigated. Stagnation points on the body surface lead
generically to the formation of cusps, whose number depends on the flow
configuration and/or the presence of nearby \tcb{boundaries}.
\end{abstract}          
%
%
%
%
%


\maketitle

\section{Introduction}\label{sec:Intro}

Fluid--body interactions at microscopic length scales or in highly
viscous flows are of common interest in physics, biology, and
engineering. Of fundamental interest in such systems is the
distribution of force per unit area, or traction, on an immersed body
as it moves through the fluid. \tcb{Some of the most exciting recent
investigations of fluid--body interactions have appeared in the
biophysics literature.} Distributions of surface tractions play
critical roles in numerous biological processes, including
significantly affecting the structure, formation, and detachment of
biofilms
\cite{picioreanu2001two,liu2002essential,stoodley2002biofilm,donlan2002biofilms,simoes2005effect}.
The competition between surface tractions and bacterial adhesion has
numerous consequences for human health \cite{nejadnik2008bacterial},
for instance in microcirculation \cite{schmid1999biomechanics} where
fluid shear stress can cause changes in endothelial cell fluidity
\cite{butler2001shear} and cell turnover rates
\cite{davies1986turbulent}, and where it can affect platelet adhesion
\cite{ruggeri2006activation}. Plaque rupture and erosion are also
shear-stress dependent \cite{shah2002pathophysiology,groen2007plaque}.
Techniques for measuring elastic moduli of cells depend on theoretical
relations between shear stress and membrane properties
\cite{hochmuth1973measurement} which could be refined with a more
detailed understanding of stress distribution on the cell surface. If
the particle is soft, surface tractions may result in changes of the
body shape and trajectory \cite{powers2010dynamics,stone2015model},
from the dynamics of flexible filaments
\cite{hinch1976deformation,becker2001instability,ts04,kantsler2012fluctuations,lmss13,lindner2015elastic,manikantan2015buckling}
and vesicles
\cite{kraus1996fluid,abkarian2005dynamics,kantsler2006transition,noguchi2007swinging,veerapaneni2009numerical}
to the shapes of single
\cite{pozrikidis2003numerical,kaoui2009red,peng2011multiscale,li2013continuum}
and many
\cite{veerapaneni2009boundary,rahimian2010petascale,kumar2012margination,freund2014numerical,fedosov2014multiscale,kumar2015cell}
red blood cells in flow. The shapes of cells and membranes in flow are
determined by a balance between viscous tractions and elastic membrane
forces \cite{pozrikidis2003numerical} including the lifting of vesicles
near walls by shear flows
\cite{sukumaran2001influence,abkarian2005dynamics,zhao2011dynamics}.

Viscous tractions also enable self-propulsion. Microorganisms use
flagellar undulations, ciliary wave propagation, and other mechanisms
to generate tractions which propel them through the fluid
\cite{lp09}. A~common modeling approach in ciliary locomotion and mucus
transport alike is to treat the layer of ciliary actuation as a smooth
traction surface
\cite{keller1975traction,sleigh1988propulsion,kanevsky2010modeling}.
Distribution of traction is important for efficient locomotion
\cite{michelin2010efficiency} as it enters locally into the description
of the required power \cite{childress1981mechanics,lp09}. The
presence of walls can reduce the effectiveness of a given ciliary
traction, resulting in slower swimming speeds of \emph{Paramecia} near
surfaces \cite{jana2012paramecium,zhu2013low,li2015swimming}.
Placement of actuation precisely where fluid stress is extremal in the
mobility problem can be used to optimize swimming speeds in viscous
flows \cite{spagnolie2010jet}.

\tcb{Other areas in which fluid--body interactions are still actively
investigated are of a more non-biological character.} \tcb{For bodies
composed of some types of materials,} surface tractions may result in
local material loss, leading to the selection of certain eroding body
geometries \cite{ristroph2012,moore2013self,mac2015shape}. At
large length and velocity scales, mass transfer can be influenced
heavily by background flows in dissolution
\cite{garner1954mass,hanratty1981stability,daccord1987fractal,rycroft2016asymmetric}
and melting \cite{hao2002heat,kumar2010heat}. At smaller length
scales there are important applications in chemical and industrial
engineering, and geophysics, from mass transfer from one or two spheres,
cylinders, or bodies of revolution
\cite{garner1958mass,linton1960transfer,steinberger1960mass,grafton1963prediction,lochiel1964mass,aminzadeh1974mass},
to more general shapes and finer-scale details
\cite{jeschke2002dissolution,colombani2008measurement,mbogoro2011intrinsic}.
Among other pharmaceutical applications are hydrodynamic manipulation
of dissolution for drug-delivery purposes
\cite{nelson1975convective,grijseels1981hydrodynamic,pillay1999unconventional,missel2004reexamination,dokoumetzidis2006century,d2006evaluation,bai2009hydrodynamic}.

There have been numerous efforts to determine surface tractions in
Stokes (zero Reynolds number\footnote{The Reynolds number is defined
by $\mbox{Re}=\rho U L/\mu $, where $\rho $ is the fluid density,
$\mu $ is the viscosity, and $U$ and $L$ are characteristic velocity and
length scales in the problem; \tcb{zero Reynolds number flows result
from the mathematical limit $\mbox{Re}\to 0$}.}) flows. Methods based
on integral equations are attractive because of the associated dimension
reduction. Such techniques can be classified in several ways: according
to whether a first- or second-kind Fredholm integral equation
\cite{stakgold2011green} is solved; whether or not the nullspace of the
integral operator is trivial; by the type of discontinuity encountered
in the integrands; whether the traction vectors are obtained directly
from the linear system or indirectly from a secondary variable; and
finally according to simplicity of derivation and implementation.

The earliest relevant works used first-kind integral formulations; in
many cases these methods enjoyed some success despite the ill-posedness
of the continuous problem and associated conditioning issues in
discretizations \cite{youngren1975stokes}. Second-kind
formulations emerged later, likely because of the impossibility of
representing an arbitrary Stokes flow using the double-layer potential
alone. An important step forward occurred in 1987 when Power and Miranda
used a double-layer potential together with a \emph{completion flow} to
represent the velocity exterior to a particle \cite{pm87}. The
completion flow is capable of exerting an arbitrary finite force and
torque on the particle; the inclusion of this flow remedies the rank
deficiency of the double-layer integral operator. \tcb{In Power and
Miranda's formulation, the completion flow consists of a point force and
point torque located at the particle centroid; a recent alternative
strategy is to use a uniform surface distribution of point forces
\cite{2016arXiv160607428C}.} We follow
\cite{KeavenyShelley2011} in referring to the completion flow method,
as generalized and refined in \cite{kk91}, as the \emph{completed
double-layer boundary integral equation} or CDLBIE.

As a method for determining fluid velocities and the net force and
torque on suspended particles, the CDLBIE has been very successful;
\tcb{moreover, this formulation can be used to evaluate the bulk fluid
flow accurately even near particle boundaries
\cite{afKlinteberg16}}. Where surface tractions are required,
several methods have been proposed. Kim and Karrila used a Riesz
reciprocal theorem argument and the CDLBIE to derive a second-kind
integral equation for the tractions on rigid particles in a quiescent
fluid \cite{kk91}. This equation is known as the \emph{completed
traction boundary integral equation} or CTBIE. If one accepts the
restriction to rigid motions, the CTBIE has all of the properties
identified as being desirable above. Moreover, the derivation of the
CTBIE using the Riesz lemma is admirably concise
\cite{KeavenyShelley2011}. Liron and Barta obtained another second-kind
traction integral equation by differentiating the single-layer velocity
field \cite{liron1992motion}, although this leads to a
noninvertible integral operator which then needs to be completed after
the fact. Ingber and Mondy gave a third traction integral equation
\cite{Ingber1993direct}; their formulation involves hypersingular
integrals, which the authors mitigate by presenting a regularization
procedure. In most of the pointwise traction computations carried out
in the literature the fluid is assumed to be at rest at infinity; one
exception is an effort by Pozrikidis who studied shear flow over a
protuberance on a wall using a first-kind integral equation
\cite{pozrikidis97}.

In this work we present an integral equation for determining the surface
tractions on a rigid body immersed in a Stokes flow which incorporates
the effect of a background flow and/or the presence of a nearby wall.
The Lorentz reciprocal theorem is used to link the surface tractions on
the body to integrals involving the background velocity and stress
fields on an imaginary bounding sphere (or hemisphere for wall-bounded
flows). The formulation reduces to an especially simple form when the
background flow is linear and the radius of the bounding surface is
taken to infinity. The method requires the velocity and stress fields
associated with numerous fundamental singularity solutions, and various
integrals including these fields, which we provide for free-space and
wall-bounded domains. The inclusion of a background flow is shown to
alter only the right-hand side of the system whereas the inclusion of
a no-slip plane wall affects only the integral operator. Two sample
applications are discussed. First, the method is used to compute the
pointwise traction on a spheroidal body as it \tcb{moves} near a
no-slip wall, which provides a more detailed understanding of
well-studied glancing and reversing trajectories \tcb{observed in the
sedimentation of particles near surfaces}. Second, we explore a new
problem, the erosion of bodies by a \tcb{Stokes} flow, in which the
surface is ablated at a rate proportional to the local viscous shear
stress. Simulations and analytical estimates suggest that a spherical
body in a uniform flow first reduces to nearly (but not exactly) the
drag minimizing profile of a body in a Stokes flow and then vanishes in
finite time. The shape dynamics of an eroding body in a shear flow and
near a wall are also investigated. Stagnation points of the flow on the
body lead generically to the formation of cusps, whose number depends
on the flow configuration and/or the presence of a nearby wall.

The paper is organized as follows. After some preliminaries in
\S \ref{sec:background}, the new integral equations are derived in
\S \ref{sec:derivation} and a discretization strategy suitable for
treating particles of arbitrary shape is described in
\S \ref{sec:numeric}. Applications are the topic of
\S \ref{sec:applications}; namely, the traction fields on sedimenting
particles near no-slip walls are computed and discussed in
\S \ref{sec:arrows}, and simulations of eroding particles in uniform and
shear flows are performed in section \S \ref{sec:erosion}, with an
accompanying analysis. We conclude with a discussion in
\S \ref{sec:conclusion}.

\section{Preliminaries}\label{sec:background}

The Stokes equations of viscous flow, valid for vanishingly small
Reynolds numbers, are given by
%
\begin{gather}
\label{eq:Stokes}
-\nabla p + \mu \nabla^{2} \bm{u}= \bm{0}\\
\nabla \cdot \bm{u} = 0,
\label{eq:unforced_stokes_pdes}
\end{gather}
where $\bm{u}$ is the velocity field, $p$ is the pressure, and
$\mu $ is the viscosity \cite{Batchelor67}. The linearity of the
Stokes equations enables the use of powerful techniques associated with
Green's functions. In this section we discuss three of these singular
solutions and collect some useful properties, including two formulas
concerning the \emph{double-layer potential} or distribution of
stresslet velocities.

In any geometry, the \emph{Stokeslet} or \emph{Lorentzlet}\footnote{The term \emph{Lorentzlet} is not in current use, but this is a little
unfair to H.A. Lorentz, who employed this tensor as early as 1896
despite some initial misgivings about the unbounded velocity near the
source point; see the exposition by Kuiken \cite{kuiken1996ha} on
Lorentz's original paper \cite{lorentz1896}.} is defined as the
solution of
%
\begin{gather}
-\nabla p + \mu \nabla^{2} \bm{u} = -\delta (\bm{x}-\bm{y})\bm{f},
\\
\nabla \cdot \bm{u} = 0,
\label{eq:lorentzlet_forced_defn}
\end{gather}
where $\delta (\bm{x}-\bm{y})\bm{f}$ is a point force, $\bm{y}$ is the
\emph{source point}, and all derivatives are with respect to
$\bm{x}$. As the system is linear, the velocity, pressure, and stress
fields associated with the point force are obtained by contracting
tensors $G$, $G^{P}$, and $G^{STR}$, respectively, against constant
multiples of $\bm{f}$:
%
\begin{gather}
u_{i}(\bm{x}) = \frac{1}{8\pi \mu }G_{ij}(\bm{x},\bm{y})f_{j},
\\
p(\bm{x}) = \frac{1}{8\pi }G^{P}_{j}(\bm{x},\bm{y})f_{j},
\\
\sigma_{ik}(\bm{x}) = \frac{1}{8\pi }G^{STR}_{ijk}(\bm{x},\bm{y})f
_{j}.
\label{eq:lorentzlet_fields}
\end{gather}
If the fluid is unbounded these tensors take simple forms; they are
provided in Table \ref{tbl:free_singularities} (see
\cite{Pozrikidis92}). Fundamental \emph{rotlet} tensor solutions $R$,
$R^{P}$, $R^{STR}$, associated instead with a point torque in the fluid,
have associated velocity $R_{ij}L_{j}/(8\pi \mu )$, pressure
$R^{P}_{j}L_{j}/(8\pi )$, and stress $R^{STR}_{ijk}L_{j}/(8\pi )$ (see
Table \ref{tbl:free_singularities}). Finally, a~\emph{stresslet}
solution, so named because its velocity field in free space arises upon
two contractions with the Stokeslet stress tensor $G_{ijk}^{STR}$, is
a Stokes flow solution with velocity, pressure, and stress given by
%
\begin{gather}
u_{i}(\bm{x}) = \frac{1}{8\pi }T_{ijk}(\bm{x},\bm{y})s_{jk},
\\
p(\bm{x}) = \frac{\mu }{8\pi }T^{P}_{jk}(\bm{x},\bm{y})s_{jk},
\\
\sigma_{im}(\bm{x}) = \frac{\mu }{8\pi }T^{STR}_{ijkm}(\bm{x},\bm{y})s
_{jk},
\end{gather}
such that $T_{ijk} = G_{ijk}^{STR}$. The units of $s_{jk}$ are area
times velocity; in this paper $s_{jk} = \psi_{j}n_{k}$ always has rank
one. Algebraic expressions for all of these tensors appear in Table \ref{tbl:free_singularities} for an unbounded fluid. Half-space
solutions may be obtained from the free-space versions by the addition
of image systems which result in a no-slip plane wall; these image
systems may be derived using the reflection principle of Lorentz, as
described in \ref{appA}. The half-space tensors for the velocity,
pressure, and stress fields above are included as Table \ref{tbl:wall_singularities}.

\begin{table}
\caption{Formulas for the Stokeslet, rotlet, and stresslet in an unbounded
fluid, each with an associated velocity, pressure, and stress field. Here
$\bm{X} = \bm{x} - \bm{y}$, where $\bm{x}$ is the observation point and $\bm{y}$ is the source point, $R=|\bm{X}|$, $\delta_{ij}$ denotes the identity
operator and $\epsilon_{ijk}$ is the alternating tensor or Levi-Civita symbol.}
\label{tbl:free_singularities}
{\footnotesize
\begin{tabular*}{\textwidth}{lclcl}
\hline
Velocity&\qquad&Pressure&\qquad&Stress  \\
\hline
$\displaystyle G_{ij} = \frac{\delta _{ij}}{R} + \frac{ X_{i} X_{j}}{R^{3}}$
&\;\;\;& $\displaystyle G^{P}_{j} = 2\frac{ X_{j}}{R^{3}}$
&\;\;\;& $\displaystyle G_{ijm}^{STR} = -6\frac{X_{i}X_{j}X_{m}}{R^{5}}$
\hfill\textcolor{white}{\framebox(0,16){}}
\\[2mm]
$\displaystyle R_{ij} = \frac{\epsilon _{ijp} X_{p}}{R^{3}} $
&& $\displaystyle R^{P}_{j} = 0 $
&& $\displaystyle R_{ijm}^{STR} = -3\frac{\epsilon _{ijp}X_{m}X_{p} + \epsilon _{mjp}X_{i}X_{p}}{R^{5}} $
\\[2mm]
$\displaystyle T_{ijk} = -6\frac{ X_{i} X_{j} X_{k}}{R^{5}} $
&& $\displaystyle T^{P}_{jk} = 4\frac{\delta _{jk}}{R^{3}} - 12\frac{ X_{j} X_{k}}{R^{5}} $
&& $\displaystyle T_{ijkm}^{STR} = -4\frac{\delta _{im}\delta _{jk}}{R^{3}} - 6\frac{\delta _{jm} X_{i} X_{k}+\delta _{km} X_{i} X_{j}+\delta _{ij} X_{k} X_{m}+\delta _{ik} X_{j} X_{m}}{R^{5}} +60\frac{X_{i} X_{j} X_{k} X_{m}}{R^{7}}$
\;\vspace{.4em}\;\\
\hline
\end{tabular*}
}
\end{table}
A \emph{double-layer potential} is a velocity field formed from a
distribution of stresslets over a closed surface,
%
\begin{equation}
u^{DL}_{i}(\bm{x}) = \frac{1}{8\pi }\int_{D} T_{ijk}(\bm{x},\bm{y})
\psi_{j}(y)n_{k}(\bm{y})\;dS_{\bm{y}}.
\label{eq:double_layer_defn}\end{equation}
Here $\bm{\psi }$ is an arbitrary density function with units of
velocity, and the normal vector, as throughout the paper, points out of
the body surface into the fluid domain. The velocity field \eqref{eq:double_layer_defn} is smooth in the interior and exterior of
$D$ but has a bounded jump discontinuity across $D$.

In the remainder of this section we will demonstrate two useful
identities concerning double-layer potentials. The first of these gives
the tensor resulting when $\psi_{j}$ is omitted from the integrand in \reftext{\eqref{eq:double_layer_defn}}. The value depends on whether the
observation point $\bm{x}$ lies within, exterior to, or on the
distribution boundary $D$. By the definition of the stresslet and \reftext{\eqref{eq:lorentzlet_forced_defn}}, we have $(\partial /\partial x_{k})
T_{ijk}(\bm{x},\bm{y})/8\pi = -\delta_{ij}\delta (\bm{x}-\bm{y})$. The
divergence theorem then gives the identity we seek:
%
\begin{align}
\frac{1}{8\pi }\int_{D} T_{ijk}(\bm{x},\bm{y})n_{k}(\bm{y})\;dS_{
\bm{y}}
=-\delta_{ij}\int_{V}\delta (\bm{x}-\bm{y})\;dV_{\bm{y}}
=
\begin{cases}
0
&\bm{x}\text{ exterior to }D
\\
- \delta_{ij}
&\bm{x}\text{ interior to }D
\\
-\delta_{ij}/2
&\bm{x}\text{ on }D.
\end{cases}
\label{eq:jump_equation}
\end{align}

The second formula we wish to establish gives the limiting value of the
double-layer potential as the observation point $\bm{x}^{e}$ approaches
a point $\bm{x}$ on the distribution surface $D$ from the exterior side
of $D$. The argument uses first the exterior case and then the boundary
case in \reftext{\eqref{eq:jump_equation}}:
%
\begin{align}
\begin{split}
\lim_{\bm{x}^{e}\to \bm{x}} \frac{1}{8\pi }\int_{D} T_{ijk}(\bm{x}
^{e},\bm{y})n_{k}(\bm{y})\psi_{j}(\bm{y})\;dS_{\bm{y}} - 0
&=
\lim_{\bm{x}^{e}\to \bm{x}} \frac{1}{8\pi }\int_{D} T_{ijk}(\bm{x}
^{e},\bm{y})n_{k}(\bm{y})\left[ \psi_{j}(\bm{y})-\psi_{j}(\bm{x})\right]
\;dS_{\bm{y}}
\\
&= \frac{1}{8\pi }\int_{D} T_{ijk}(\bm{x},\bm{y})n_{k}(\bm{y})\left[ \psi
_{j}(\bm{y})-\psi_{j}(\bm{x})\right] \;dS_{\bm{y}}
\\
&= \frac{1}{8\pi }\int_{D} T_{ijk}(\bm{x},\bm{y})n_{k}(\bm{y})\psi
_{j}(\bm{y})\;dS_{\bm{y}} + \frac{1}{2}\psi_{i}(\bm{x}).
\end{split}
\label{eq:limiting_double_layer}
\end{align}
The evaluation of the limit is legitimate because of the regularity
provided by the extra zero factor $(\psi_{j}(\bm{y}) - \psi_{j}(
\bm{x}))$; for more details see the book by Pozrikidis
\cite{Pozrikidis92}. Both equations \reftext{\eqref{eq:jump_equation}} and \reftext{\eqref{eq:limiting_double_layer}} will be employed below.

\section{Generalized traction integral equations}\label{sec:derivation}

We now show how the argument presented in
\cite{KeavenyShelley2011} can be extended to solve the resistance
problem with nonzero background flow and/or a no-slip plane wall. We
find that the background flow affects the right-hand side of the system
but not the integral operator, whereas the presence of a plane wall
affects the integral operator but not the right-hand side. The
corresponding mobility problems are discussed later.

\subsection{A background flow in free space}

Let $D$ denote the boundary of a rigid particle immersed in an unbounded
fluid with centroid $\bm{Y}$, and let $S$ denote a sphere of large
radius centered at the origin and containing the particle. The Lorentz
reciprocal theorem states that two Stokes solutions with velocity and
stress fields $(\bm{u}, \bm{\sigma })$ and $(\bm{u}', \bm{\sigma }')$
satisfy \cite{lorentz1907,Pozrikidis92}:
%
\begin{gather}
\bm{\nabla} \cdot (\bm{u}' \cdot \bm{\sigma }-\bm{u} \cdot \bm{\sigma }')=
0.
\end{gather}
Integrating the above over a volume bounded by $D$ and $S$, and applying
the divergence theorem, we have
%
\begin{equation}
\label{eq:Lorentz}
\langle \bm{u}', \bm{f}\rangle_{D} + \langle \bm{u}', \bm{f}\rangle
_{S} = \langle \bm{u}, \bm{f}'\rangle_{D} + \langle \bm{u}, \bm{f}'
\rangle_{S},
\end{equation}
where $\bm{f}$ and $\bm{f}'$ are the surface tractions exerted by the
flows $\bm{u}$ and $\bm{u}'$, respectively, and $\langle \bm{u}',
\bm{f} \rangle_{D} = \int_{D} \bm{u}'\cdot \bm{f}\,dS$. The first
solution $(\bm{u},\bm{\sigma })$ will solve the resistance problem of
principal interest: the flow $\bm{u}$ arises when the body moves with
specified velocity $\bm{U}$ and rotation vector $\bm{\Omega }$ in a
fluid with specified background flow/pressure fields $\bm{u}^{\infty
},\, p^{\infty }$. It is the surface tractions $\bm{f}$ associated with
this flow that we ultimately seek. Boundary conditions for $\bm{u}$ are
given by the no-slip boundary condition on $D$, $\bm{u} = \bm{U} +
\bm{\Omega }\times (\bm{x} - \bm{Y})$, and by decay towards the
background flow field at infinity, $|\bm{u} - \bm{u}^{\infty }| =
\mathcal{O}\left( 1/|\bm{x}-\bm{Y}|\right) $ as $|\bm{x} - \bm{Y}|
\to \infty $. The power of the reciprocal theorem becomes apparent upon
the selection of the second, auxiliary solution. Let $\bm{u}'$ denote
a velocity field consisting of the sum of a double-layer potential and
a \emph{completion flow}; specifically, for $\bm{x}$ in the exterior of
the body we define
%
\begin{equation}
u_{i}'(\bm{x}) = \frac{1}{8\pi }\int_{D} T_{ijk}(\bm{x},\bm{y})n_{k}(
\bm{y})\psi_{j}(\bm{y})\;dS_{\bm{y}} + \frac{c}{8\pi }\int_{D} C_{ij}(
\bm{x},\bm{y})\psi_{j}(\bm{y})\;dS_{\bm{y}},
\label{eq:exterior_formula_tc}
\end{equation}
where $\bm{n}$ is the normal vector pointing into the fluid from the
particle, $T_{ijk}$ is the free-space stresslet, and $\bm{\psi }$ is an
arbitrary smooth vector field on $D$ with units of velocity. The
proportionality constant $c$ and the tensor $C_{ij}$ both have units of
inverse length. The completion flow is required to be regular both on
and exterior to $D$, and moreover it must exert a total force and torque
$\bm{F} = c\mu \int_{D} \bm{\psi }(\bm{y})\;dS_{\bm{y}}$ and
$\bm{L} = c\mu \int_{D}(\bm{y} - \bm{Y})\times \bm{\psi }\;dS_{\bm{y}}$
on the particle. The completion flow can be specified in several ways;
later we will consider a specific choice but for now we leave the
expressions general for the benefit of others who may wish to use a
different completion flow. The purpose of including the completion flow
is to complete the range of the integral operator, since the double
layer potential alone is incapable of exerting a net force or torque on
the particle \cite{Pozrikidis92}.

The next step is to write all four integrals in
\reftext{\eqref{eq:Lorentz}} as inner products against $\bm{\psi }$.
Two of the terms have been discussed elsewhere
\cite{KeavenyShelley2011,kk91}; we include the arguments here for
completeness. For the term $\langle \bm{u}', \bm{f}\rangle_{D}$, we
need an expression for the double layer potential in the limit where
the observation point $\bm{x}^{e}$ approaches a boundary point
$\bm{x}\in D$ from the exterior,
i.e.~\reftext{\eqref{eq:limiting_double_layer}}. Using this we obtain
%
\begin{align}
\begin{split}
\langle \bm{u}', \bm{f}\rangle_{D}
&= \int_{D} f_{i}(\bm{x}) \left[ \frac{1}{2}
\psi_{i}(\bm{x}) + \frac{1}{8\pi }\int_{D} T_{ijk}(\bm{x},\bm{y})n
_{k}(\bm{y})\psi_{j}(\bm{y})\;dS_{y} + \frac{c}{8\pi }\int_{D} \psi
_{j}(\bm{y}) C_{ij}(\bm{x},\bm{y})\;dS_{\bm{y}}\right] \;dS_{\bm{x}}
\label{eq:psi_evaluated_on_particle_surface}
\\
&=\int_{D} f_{i}(\bm{x}) \int_{D}\psi_{j}(\bm{y})\left[ \frac{1}{2}
\delta_{ij}\delta (\bm{x}-\bm{y}) + \frac{1}{8\pi }T_{ijk}(\bm{x},
\bm{y})n_{k}(\bm{y}) + \frac{c}{8\pi }C_{ij}(\bm{x},\bm{y})\right] dS
_{\bm{y}}\;dS_{\bm{x}}
\\
&=\int_{D} \psi_{j}(\bm{y}) \left[ \frac{1}{2}f_{j}(\bm{y}) + \frac{n
_{k}(\bm{y})}{8\pi }\int_{D}T_{ijk}(\bm{x},\bm{y})f_{i}(\bm{x})dS_{
\bm{x}} + \frac{c}{8\pi }\int_{D} C_{ij}(\bm{x},\bm{y})f_{i}(\bm{x})dS
_{\bm{x}}\right] dS_{\bm{y}}.
\end{split}
\end{align}
For the term $\langle \bm{u}, \bm{f}'\rangle_{D}$, we use the fact that
$\bm{u}$ is a rigid body motion, along with the relations between
$\bm{\psi }$ and $(\bm{F},\bm{L})$ assumed above, resulting in:
%
\begin{align}
\label{eq:needs_rbm}
\begin{split}
\langle \bm{u}, \bm{f}'\rangle_{D}
&= \langle \bm{U} + \bm{\Omega }
\times (\bm{y}-\bm{Y}), \bm{f}'\rangle_{D}
= \bm{U}\cdot \bm{F}+
\bm{\Omega }\cdot \tcb{\bm{L}}
= \left\langle  c\mu \bm{U} + c\mu
\bm{\Omega }\times (\bm{y}-\bm{Y}),\bm{\psi }\right\rangle _{D}.
\end{split}
\end{align}

If there is no background flow ($\bm{u}^{\infty} =0$) then the integrals
over $S$ in \reftext{\eqref{eq:Lorentz}} both vanish as the radius of $S$ becomes
infinite and we obtain
\begin{multline}
 0 = \int_D \psi_j(\bm y) \left[c\mu\left( U_j + (\bm \Omega\times (\bm x - \bm Y))_j\right)   -\frac{1}{2}f_j(\bm y) - \frac{1}{8\pi}n_k(\bm y)\int_DT_{ijk}(\bm x,\bm y)f_i(\bm x)dS_{\bm x}\right. \\
  - \left.\frac{c}{8\pi}\int_D C_{ij}(\bm x,\bm y)f_i(\bm x)dS_{\bm x}\right]dS_{\bm y}\label{eq:previouslystudied_yes_psi},
\end{multline}
%
which, because $\bm{\psi }$ is arbitrary, leads to the desired integral
equation for $\bm{f}$. In our notation this is
%
\begin{equation}
\frac{1}{2}f_{j}(\bm{y}) + \frac{1}{8\pi }n_{k}(\bm{y})\int_{D}T_{ijk}(
\bm{x},\bm{y})f_{i}(\bm{x})dS_{\bm{x}} + \frac{c}{8\pi }\int_{D} C
_{ij}(\bm{x},\bm{y})f_{i}(\bm{x})dS_{\bm{x}} = c\mu \left(  U_{j} + (
\bm{\Omega }\times (\bm{x} - \bm{Y}))_{j}\right)
\label{eq:previouslystudied_no_psi},
\end{equation}
and this holds for each $\bm{y}\in D$ and for each $j=1,2,3$. This is
equivalent to equation (20) in \cite{KeavenyShelley2011}. The
previously published equation is identical despite our inclusion of a
minus sign in the stresslet, as is more conventional; the sign change
is effectively absorbed by the freedom of choosing the completion flow.

If there is in fact a background flow, then $\langle \bm{u}', \bm{f}
\rangle_{S} $ and $\langle \bm{u}, \bm{f}'\rangle_{S}$ do not vanish as
the radius of the bounding sphere $S$ tends to infinity. We must
therefore write these also in the form $\langle \bm{\psi }, \cdot
\rangle_{D}$. For the first we have
%
\begin{align}
\begin{split}
\langle \bm{u}', \bm{f}\rangle_{S}
&= \langle \bm{u}', \bm{f}^{\infty
}\rangle_{S}
\\
&= \int_{S} f_{i}^{\infty }(\bm{x}) \int_{D} \psi_{j}(\bm{y})\left[  \frac{1}{8
\pi }T_{ijk}(\bm{x},\bm{y})n_{k}(\bm{y}) + \frac{c}{8\pi }C_{ij}(
\bm{x},\bm{y})\right] dS_{\bm{y}}dS_{\bm{x}}
\\
&= \int_{D} \psi_{j}(\bm{y}) \int_{S} f_{i}^{\infty }(\bm{x})\left[  \frac{1}{8
\pi }T_{ijk}(\bm{x},\bm{y})n_{k}(\bm{y}) + \frac{c}{8\pi }C_{ij}(
\bm{x},\bm{y})\right] dS_{\bm{x}}dS_{\bm{y}}.
\end{split}
\end{align}
The remaining integral involves the traction field $\bm{f}'$ of the
flow arising from $\bm{\psi }$ at the inner surface of the bounding
sphere~$S$, so the stress fields of the stresslet and of the completion
flow are required. The tensor $T^{STR}_{ijkm}$ generating the stress
field corresponding to the stresslet velocity field was given in
\reftext{Table~\ref{tbl:free_singularities}}. Denote by
$C_{ijk}^{STR}(\bm{x},\bm{y})$ the kernel for the stress field of the
completion flow, which we leave general for the moment. Then we may
proceed to write
%
\begin{align}
\begin{split}
\langle \bm{u}, \bm{f}'\rangle_{S}
&= \langle \bm{u}^{\infty },
\bm{f}'\rangle_{S}
\\
&= \int_{S} u_{i}^{\infty }(\bm{x})\sigma_{im}'(\bm{x}){\hat{n}}_{m}(
\bm{x})\;dS_{\bm{x}}
\\
&= \int_{S} u_{i}^{\infty }(\bm{x}){\hat{n}}_{m}(\bm{x})\int_{D}\psi
_{j}(\bm{y})\left[ \frac{\mu }{8\pi }T^{STR}_{ijkm}(\bm{x},\bm{y})n
_{k}(\bm{y}) + \frac{c\mu }{8\pi }C^{STR}_{ijm}(\bm{x},\bm{y}) \right] dS
_{\bm{y}}dS_{\bm{x}}
\\
&= \int_{D} \psi_{j}(\bm{y})\int_{S} u_{i}^{\infty }(\bm{x}){\hat{n}}
_{m}(\bm{x})\left[ \frac{\mu }{8\pi }T^{STR}_{ijkm}(\bm{x},\bm{y})n
_{k}(\bm{y}) + \frac{c\mu }{8\pi }C^{STR}_{ijm}(\bm{x},\bm{y}) \right] dS
_{\bm{x}}dS_{\bm{y}},
\end{split}
\end{align}
where $\hat{\bm{n}}$ is an inward pointing normal vector on $S$.

Now all four terms in \reftext{\eqref{eq:Lorentz}} have been rewritten as inner
products against the arbitrary function $\bm{\psi }$. Collecting these
into a single inner product against $\bm{\psi }$ and recalling that
$\bm{\psi }$ was arbitrary, we conclude that the other argument in the
inner product must vanish identically.\footnote{More precisely, the
Riesz lemma implies that if the inner product $\langle \mathcal{F},
\psi \rangle $ vanishes for all $\psi $ in a dense subset of a Hilbert
space, then in fact $\mathcal{F} = 0$. Our Hilbert space is
$L^{2}(D)^{3}$ with inner product $\langle \bm{u},\bm{v}\rangle = \int
_{D}\bm{u} \cdot \bm{v}\;dS$ and the dense subset is the collection of
smooth functions.} This leads to an integral equation for the unknown
surface tractions $\bm{f}$ on $D$,
%
\begin{gather}
\begin{split}
&\frac{1}{2}f_{j}(\bm{y}) + \frac{1}{8\pi }n_{k}(\bm{y})\int_{D}T_{ijk}(
\bm{y}',\bm{y})f_{i}(\bm{y}')dS_{\bm{y}'}
+ \frac{c}{8\pi }\int_{D} C
_{ij}(\bm{y}',\bm{y})f_{i}(\bm{y}')dS_{\bm{y}'} = c\mu ( U_{j} +
\epsilon_{jk\ell }\Omega_{k}(y_{\ell }-Y_{\ell }))
\\
&\quad {}
+ \int_{S} u_{i}^{\infty }(\bm{x}){\hat{n}}_{m}(\bm{x})\left[ \frac{
\mu }{8\pi }T^{STR}_{ijkm}(\bm{x},\bm{y})n_{k}(\bm{y}) + \frac{c
\mu }{8\pi }C^{STR}_{ijm}(\bm{x},\bm{y}) \right] dS_{\bm{x}}
\\
&\quad {}
- \int_{S} f_{i}^{\infty }(\bm{x})\left[  \frac{1}{8\pi }T_{ijk}(\bm{x},
\bm{y})n_{k}(\bm{y}) + \frac{c}{8\pi }C_{ij}(\bm{x},\bm{y})\right] dS
_{\bm{x}},
\end{split}
\label{eq:general_nowall_shear}
\end{gather}
which holds for $j=1,2,3$ and for all $\bm{y}\in D$. Comparing this to \reftext{\eqref{eq:previouslystudied_no_psi}} we see that the inclusion of a
background flow has modified only the right-hand side of the integral
equation, leaving unchanged the operator which must be inverted.

The integral equation \reftext{\eqref{eq:general_nowall_shear}} holds for any
background flow which solves the unforced Stokes equations \reftext{\eqref{eq:unforced_stokes_pdes}}, and for any completion flow. To apply
this equation for a specific background flow we must evaluate the
integrals over the bounding sphere $S$, which can have any desired
radius as long as it contains the particle. In principle this can be
done numerically, but it is more convenient if the integrals over
$S$ converge to analytically tractable expressions in the limit of large
radius. This occurs for a linear background flow, so we now take
$u_{i}^{\infty }(\bm{x}) = A_{ij}x_{j}$ and $\sigma_{ik}^{\infty }(
\bm{x}) =\mu (A_{ik}+A_{ki})$, where $A$ has zero trace to satisfy
incompressibility. At the same time we assume a completion flow with the
kernel
%
\begin{equation}
C_{ij}(\bm{x}, \bm{y}) =G_{ij}(\bm{x},\bm{z}(\bm{y}))
+ \epsilon_{m
\ell j} R_{im}(\bm{x},\bm{z}(\bm{y}))(y_{\ell }-z_{\ell }(\bm{y})),
\label{eq:free_completion_flow}
\end{equation}
where $G_{ij}$ and $R_{ij}$ are given in
\reftext{Table~\ref{tbl:free_singularities}} and $\bm{z}: D\to
\operatorname{Int}D$ is a map from the surface to the interior of the
particle; in the case that $\bm{z}\equiv \bm{Y}$ (a Stokeslet and
rotlet are placed at the particle centroid) this is the completion flow
of Power and Miranda~\cite{pm87}. The kernel for the stress field of
this flow is
%
\begin{equation}
C^{STR}_{ijk}(\bm{x},\bm{y})=\mu G^{STR}_{ijk}(\bm{x}-\bm{z}(\bm{y}))
+\mu \epsilon_{m\ell j} R^{STR}_{imk}(\bm{x},\bm{z}(\bm{y}))(y_{
\ell }-z_{\ell }(\bm{y})),
\end{equation}
where $G^{STR}$ and $R^{STR}$ are given in
\reftext{Table~\ref{tbl:free_singularities}}. The integrals on the
right-hand side of \reftext{\eqref{eq:general_nowall_shear}} are
evaluated in \ref{appB}, resulting in the final integral equation for
tractions on a rigid body moving with velocity $\bm{U}+\bm{\Omega
}\times (\bm{x}-\bm{Y})$ in an infinite fluid with background flow
$\bm{u}^{\infty }(\bm{x}) = A\bm{x}$:
%
\begin{equation}
\begin{split}
&\frac{1}{2}f_{j}(\bm{y}) + \frac{1}{8\pi }n_{k}(\bm{y})\int_{D}T_{ijk}(
\bm{y}',\bm{y})f_{i}(\bm{y}')dS_{\bm{y}'}
+ \frac{c}{8\pi }\int_{D} C
_{ij}(\bm{y}',\bm{y})f_{i}(\bm{y}')dS_{\bm{y}'}
\\
&\quad {}
= c\mu \left( U_{j} + \epsilon_{jk\ell }\Omega_{k}(y_{\ell }-Y_{
\ell })\right)  - \mu (A_{jk}+A_{kj})n_{k}(\bm{y}) + \frac{c\mu }{2}(A
_{jk}-A_{kj})y_{k} +\frac{c\mu }{2}(A_{jk}+A_{kj})z_{k}(\bm{y}).
\end{split}
\label{eq:linear_free}
\end{equation}
Note that the identity \reftext{\eqref{eq:jump_equation}} and the antisymmetry
property $T_{ijk}(\bm{x},\bm{y}) = - T_{ijk}(\bm{y},\bm{x})$ of the
free-space stresslet lead to a singularity subtraction which gives a
version better suited for discretization:
%
\begin{gather}
\begin{split}
&\frac{1}{8\pi }\int_{D}T_{ijk}(\bm{y}',\bm{y})\Big(f_{i}(\bm{y}')n
_{k}(\bm{y}) + f_{i}(\bm{y})n_{k}(\bm{y}')\Big)dS_{\bm{y}'}
+ \frac{c}{8
\pi }\int_{D} C_{ij}(\bm{y}',\bm{y})f_{i}(\bm{y}')dS_{\bm{y}'}
\\
&\quad {}
= c\mu \left( U_{j} + \epsilon_{jk\ell }\Omega_{k}(y_{\ell }-Y_{
\ell })\right)  - \mu (A_{jk}+A_{kj})n_{k}(\bm{y}) + \frac{c\mu }{2}(A
_{jk}-A_{kj})y_{k} +\frac{c\mu }{2}(A_{jk}+A_{kj})z_{k}(\bm{y}).
\label{eq:subtracted_linear_free}
\end{split}
\end{gather}
The subtracted singularity has the regularity of a bounded jump
discontinuity \cite{KeavenyShelley2011}.

\subsection{Near a no-slip wall}

A no-slip wall at $\{x_{3} = 0\}$ can be accounted for with only a few
modifications to the preceding argument. We begin by replacing the
free-space stresslet $T_{ijk}(\bm{x},\bm{y})$ in \reftext{\eqref{eq:exterior_formula_tc}} with its half-space counterpart,
$T_{ijk}^{\text{half}}(\bm{x},\bm{y})$, which is derived using the
Lorentz reflection procedure (see \ref{appA}) and given in \reftext{Table~\ref{tbl:wall_singularities}}. The completion flow in \reftext{\eqref{eq:exterior_formula_tc}} is subject to the additional requirement
of vanishing on the wall, but is allowed to become singular below the
wall. We then apply the reciprocal theorem on the volume external to the
particle, above the wall, and below a large hemisphere $H$ which is
centered at $(Y_{1}, Y_{2}, 0)$. Both flows vanish on the wall, and if
there is no background flow the integrals over $H$ decay as the radius
of $H$ increases, yielding \reftext{\eqref{eq:previouslystudied_no_psi}} but with
the free-space singularities replaced by their wall-bounded
counterparts; the integral equation is then
%
\begin{gather}
\label{eq:quiet_wall_ie}
\frac{1}{2}f_{j}(\bm{y}) + \frac{1}{8\pi }n_{k}(\bm{y})\int_{D}T^{
\text{half}}_{ijk}(\bm{x},\bm{y})f_{i}(\bm{x})dS_{\bm{x}} + \frac{c}{8
\pi }\int_{D} C^{\text{half}}_{ij}(\bm{x},\bm{y})f_{i}(\bm{x})dS
_{\bm{x}} = c\mu \left( U_{j} + (\bm{\Omega }\times (\bm{x} -
\bm{Y}))_{j}\right) .
\end{gather}
With a nonzero background flow the right-hand side includes integrals
over $H$:
\begin{multline}
\label{eq:background_wall_general_ie}
\frac{1}{2}f_{j}(\bm{y}) + \frac{1}{8\pi }n_{k}(\bm{y})\int_{D}T^{
\text{half}}_{ijk}(\bm{x},\bm{y})f_{i}(\bm{x})dS_{\bm{x}} + \frac{c}{8
\pi }\int_{D} C^{\text{half}}_{ij}(\bm{x},\bm{y})f_{i}(\bm{x})dS
_{\bm{x}}
\\
= c\mu \left( U_{j} + (\bm{\Omega }\times (\bm{x} - \bm{Y}))_{j}\right) +
\int_{H} u_{i}^{\infty }(\bm{x}){\hat{n}}_{m}(\bm{x})\left[ \frac{
\mu }{8\pi }T^{\text{half},STR}_{ijkm}(\bm{x}-\bm{y})n_{k}(
\bm{y})+ \frac{c\mu }{8\pi }C^{\text{half},STR}_{ijm}(\bm{x},
\bm{y}) \right] dS_{\bm{x}}
\\
- \int_{H} f_{i}^{\infty }(\bm{x})\left[  \frac{1}{8\pi }T^{
\text{half}}_{ijk}(\bm{x}-\bm{y})n_{k}(\bm{y}) + \frac{c}{8
\pi }C^{\text{half}}_{ij}(\bm{x},\bm{y})\right] dS_{\bm{x}} .
\end{multline}
The only linear background flow satisfying a no-slip condition at
$\{x_{3} = 0\}$ is a shear flow. Without loss of generality we may
suppose the direction of shear flow is parallel to the $x$-direction:
$\bm{u}^{\infty }(\bm{x}) = \dot{\gamma }x_{3}\bm{\hat{x}}_{1}$. We
specialize to this case and we assume a completion flow of the form
%
\begin{align}
\begin{split}
C^{\text{half}}_{ij}(\bm{x}, \bm{y})
&=\frac{c}{8\pi }G_{ij}
^{\text{half}}(\bm{x},\bm{z}(\bm{y}))
+ \frac{c}{8\pi }
\epsilon_{m\ell j} R_{im}^{\text{half}}(\bm{x},\bm{z}(\bm{y}))(y
_{\ell }-z_{\ell }(\bm{y})),
\end{split}
\label{eq:wall_completion_flow}
\end{align}
where the half-space Stokeslet and rotlet are given in \reftext{Table~\ref{tbl:wall_singularities}}.

We then compute the integrals over $H$ in the limit of large radius.
Although the wall-bounded analogues of \reftext{\eqref{eq:int_c_str}} and \reftext{\eqref{eq:int_c_vel}} converge to different expressions than in the
free-space case, the differences are opposites and so the sum of the
four integrals is the same as above. The result is a completed traction
integral equation for the rigid motion of a single particle above a
no-slip wall in a background shear flow:
%
\begin{gather}
\begin{split}
\label{eq:background_wall_linear_ie}
&\frac{1}{2}f_{j}(\bm{y}) + \frac{1}{8\pi }n_{k}(\bm{y})\int_{D}T^{
\text{half}}_{ijk}(\bm{y}',\bm{y})f_{i}(\bm{y}')dS_{\bm{y}'}
+ \frac{c}{8
\pi }\int_{D} C^{\text{half}}_{ij}(\bm{y}',\bm{y})f_{i}(\bm{y}')dS
_{\bm{y}'}
\\
&\quad {}
= c\mu \left( U_{j} + \epsilon_{jk\ell }\Omega_{k}(y_{\ell }-Y_{
\ell })\right)  - \mu \dot{\gamma }\left( \delta_{1j}n_{3}(\bm{y}) +
\delta_{3j}n_{1}(\bm{y})\right)
\\
&\qquad {}
+ \frac{c\mu \dot{\gamma }}{2}\left( \delta
_{1j}y_{3} - \delta_{3j}y_{1}\right)  +\frac{c\mu \dot{\gamma }}{2}
\left( \delta_{1j}z_{3}(\bm{y}) + \delta_{3j}z_{1}(\bm{y})\right) .
\end{split}
\end{gather}
The right-hand side of this equation is identical to that of \reftext{\eqref{eq:linear_free}}, once we specialize to $A_{jk} = \dot{\gamma }
\delta_{1j}\delta_{3k}$.\goodbreak

We close with a comment on the singularity subtraction for the
wall-bounded operator. The only term in the expression for $T_{ijk}
^{\text{half}}(\bm{x},\bm{y})$ in \reftext{Table~\ref{tbl:wall_singularities}} which diverges as $\bm{x}\to \bm{y}$ is
precisely the free-space stresslet, so we decompose the wall-bounded
operator into bounded and diverging parts and then carry out the same
subtraction as above for the diverging part. This leads to the
subtraction
%
\begin{eqnarray}[ll]
\frac{1}{2}f_{j}(\bm{y}) + \frac{n_{k}(\bm{y})}{8\pi }\int_{D}T^{
\text{half}}_{ijk}(\bm{y}',\bm{y})f_{i}(\bm{y}')dS_{\bm{y}'}
\nonumber
\\
\quad {}
=
\frac{1}{8
\pi }\int_{D} \left[  T_{ijk}(\bm{y}',\bm{y})\Big(f_{i}(\bm{y}')n_{k}(
\bm{y}) + f_{i}(\bm{y})n_{k}(\bm{y}')\Big)
+ n_{k}(\bm{y}) T^{*}_{ijk}(
\bm{y}',\bm{y})f_{i}(\bm{y}')\right] dS_{\bm{y}'},
\end{eqnarray}
where $T^{*} = T^{\text{half}}- T$, the image singularity system
of the free-space stresslet, is regular throughout the fluid domain.
This subtraction should be applied to the equations \reftext{\eqref{eq:quiet_wall_ie}} and \reftext{\eqref{eq:background_wall_linear_ie}} before
discretization to reduce the singularity of the integrand to that of a
bounded jump discontinuity.

\section{Numerical method and verification}\label{sec:numeric}

\subsection{Discrete equations}

We now consider a simple collocation discretization of the equations
derived in the previous section. In this scheme we encode all of the
information about the particle geometry in a list of boundary points
$\bm{y}^{p}$, normal vectors $\bm{n}^{p}$ at those points, and
quadrature weights $w_{p}$ selected such that $\int_{D} \psi (\bm{x})
\,dS_{x} \approx \sum_{p=1}^{N} \psi (y^{p})w_{p}$. In particular, two
methods are used for generating the surface quadrature rules in this
paper, one which takes advantage of axisymmetric geometries and the
other applicable for particles of general shape. In the first method,
we use a spherical coordinate system and represent particle points as
$\bm{y}(\phi ,\eta ) = r(\phi ) \left( \cos (\phi ),\sin (\phi )\cos (
\eta ), \sin (\phi )\sin (\eta )\right) $ where $r(\phi )$ is a specified
profile. Gaussian quadrature is applied in the zenith angle
$\phi $ and the trapezoidal rule is used in
the azimuthal angle $\eta $. The second method is based on a triangular
mesh of the body; in particular we use \texttt{distmesh}
\cite{strang2004simple} to generate uniform triangular meshes, in which
case the quadrature nodes are the mesh vertices, the vertex normals are
the area weighted averages of adjacent face normals, and the weights are
simply one-third of the sums of areas of adjacent (flat) triangles.

With any quadrature $\{(\bm{y}^{p},\bm{n}^{p}, w_{p})\}_{p=1}^{N}$, the
free-space integral operator at the source point $\bm{y}\in D$ is
approximated by
%
\begin{align}
\begin{split}
&\frac{1}{8\pi }\int_{D}T_{ijk}(\bm{y}',\bm{y})\Big(f_{i}(\bm{y}')n
_{k}(\bm{y}) + f_{i}(\bm{y})n_{k}(\bm{y}')\Big)dS_{\bm{y}'} + \frac{c}{8
\pi }\int_{D} C_{ij}(\bm{y}',\bm{y})f_{i}(\bm{y}')dS_{\bm{y}'}
\\
&\quad {}\approx \sum_{p=1}^{N}
\left[ \frac{1}{8\pi }T_{ijk}(\bm{y}^{p},
\bm{y})\Big(f_{i}(\bm{y}^{p})n_{k}(\bm{y}) + f_{i}(\bm{y})n_{k}(
\bm{y}^{p})\Big) + \frac{c}{8\pi } C_{ij}(\bm{y}^{p},\bm{y})f_{i}(
\bm{y}^{p})\right] w_{p}.
\label{eq:quadrature}
\end{split}
\end{align}
The integral equation is required to hold at the quadrature nodes
$\bm{y} = \bm{y}^{q}$, for $q = 1,\cdots ,N$. The stresslet term (with
singularity subtraction) in the integrand of \reftext{\eqref{eq:quadrature}} has
a bounded jump discontinuity at the source point $\bm{y}$ for a smooth
surface; fortunately, in simply omitting the source point from the sum,
both quadrature schemes above retain second-order accuracy (see
\cite{pm87,KeavenyShelley2011}, and note that the local form of the jump
singularity averages to zero upon integration over a small circular
patch). That is, we determine the $3N$ unknowns $f_{j}(\bm{y}^{q})$
which solve the $3N$ equations
%
\begin{align}
\begin{split}
&\sum_{p\neq q}
\frac{1}{8\pi }T_{ijk}(\bm{y}^{p},\bm{y}^{q})\Big(f
_{i}(\bm{y}^{p})n_{k}(\bm{y}^{q}) + f_{i}(\bm{y}^{q})n_{k}(\bm{y}^{p})
\Big)w_{p} + \frac{c}{8\pi }\sum_{p=1}^{N} C_{ij}(\bm{y}^{p},\bm{y}
^{q})f_{i}(\bm{y}^{p})w_{p}
\\
&\quad {}
=c\mu \left( U_{j} + \epsilon_{jk\ell }\Omega_{k}(y^{q}_{\ell }-Y_{
\ell })\right)  - \mu (A_{jk}+A_{kj})n_{k}(\bm{y}^{q}) +
\frac{c\mu }{2}(A_{jk}-A_{kj})y^{q}_{k} +\frac{c\mu }{2}(A_{jk}+A_{kj})z
_{k}(\bm{y}^{q}),
\label{eq:discrete_free}
\end{split}
\end{align}
for $j = 1,2,3$ and $q = 1,\cdots , N$. With a wall, the discrete
equations are instead
%
\begin{align}
\begin{split}
&\sum_{p\neq q}
\frac{1}{8\pi }T_{ijk}(\bm{y}^{p},\bm{y}^{q})\Big(f
_{i}(\bm{y}^{p})n_{k}(\bm{y}^{q}) + f_{i}(\bm{y}^{q})n_{k}(\bm{y}^{p})
\Big)w_{p}
\\
&\qquad {}
+ \frac{1}{8\pi }n_{k}(\bm{y}^{q})\sum_{p=1}^{N} T^{*}_{ijk}(
\bm{y}^{p},\bm{y}^{q})f_{i}(\bm{y}^{p})w_{p} + \frac{c}{8\pi }\sum
_{p=1}^{N} C^{\text{half}}_{ij}(\bm{y}^{p},\bm{y}^{q})f_{i}(
\bm{y}^{p})w_{p}
\\
&\quad {}
= c\mu \left( U_{j} + \epsilon_{jk\ell }\Omega_{k}(y_{\ell }^{q}-Y
_{\ell })\right)  - \mu \dot{\gamma }\left( \delta_{1j}n_{3}(\bm{y}^{q})
+ \delta_{3j}n_{1}(\bm{y}^{q})\right)
\\
&\qquad {}
+ \frac{c\mu \dot{\gamma }}{2}
\left( \delta_{1j}y^{q}_{3} - \delta_{3j}y^{q}_{1}\right)  +\frac{c
\mu \dot{\gamma }}{2}\left( \delta_{1j}z_{3}(\bm{y}^{q}) + \delta_{3j}z
_{1}(\bm{y}^{q})\right) .
\end{split}
\label{eq:discrete_wall}
\end{align}

In all cases to be described we adopt the Power and Miranda completion
flow, setting $\bm{z}(\bm{y}) \equiv \bm{Y}$ with $\bm{Y}$ the particle
centroid. The discrete linear system of equations is dense and
non-normal. We solve it using the generalized minimal residual algorithm
(GMRES) \cite{ss86} without preconditioning or restarting until
a relative residual of $10^{-12}$ is reached.

\subsection{Mobility formulation}

So far, we have discussed the resistance problem where the particle
velocity is prescribed and the surface tractions as well as the net
force and torque are unknown. Solving instead the mobility problem,
where the net force and torque are imposed and the surface tractions
and the rigid-body velocity are unknown, requires only a minor
modification to~\reftext{\eqref{eq:discrete_free}}
(or~\reftext{\eqref{eq:discrete_wall}}). The terms involving the six
constants $U_{j}$ and $\Omega_{k}$ are subtracted from the right-hand
to the left-hand side and these become additional unknowns, and the
system is closed by enforcing
%
\begin{align}
\sum_{p=1}^{N} f_{j}(\bm{y}^{p}) w_{p} = F_{j},
\qquad
\sum_{p=1}^{N} \epsilon_{jk\ell } (y^{p}_{k} - Y_{k})f_{\ell }(\bm{y}
^{p}) w_{p} = L_{j},
\end{align}
for $j=1,2,3$ (recall that $\bm{F}$ and $\bm{L}$ denote the imposed net
force and torque exerted by the fluid on the particle).

\subsection{Validation}

\begin{figure}
\includegraphics[width=0.8\linewidth]{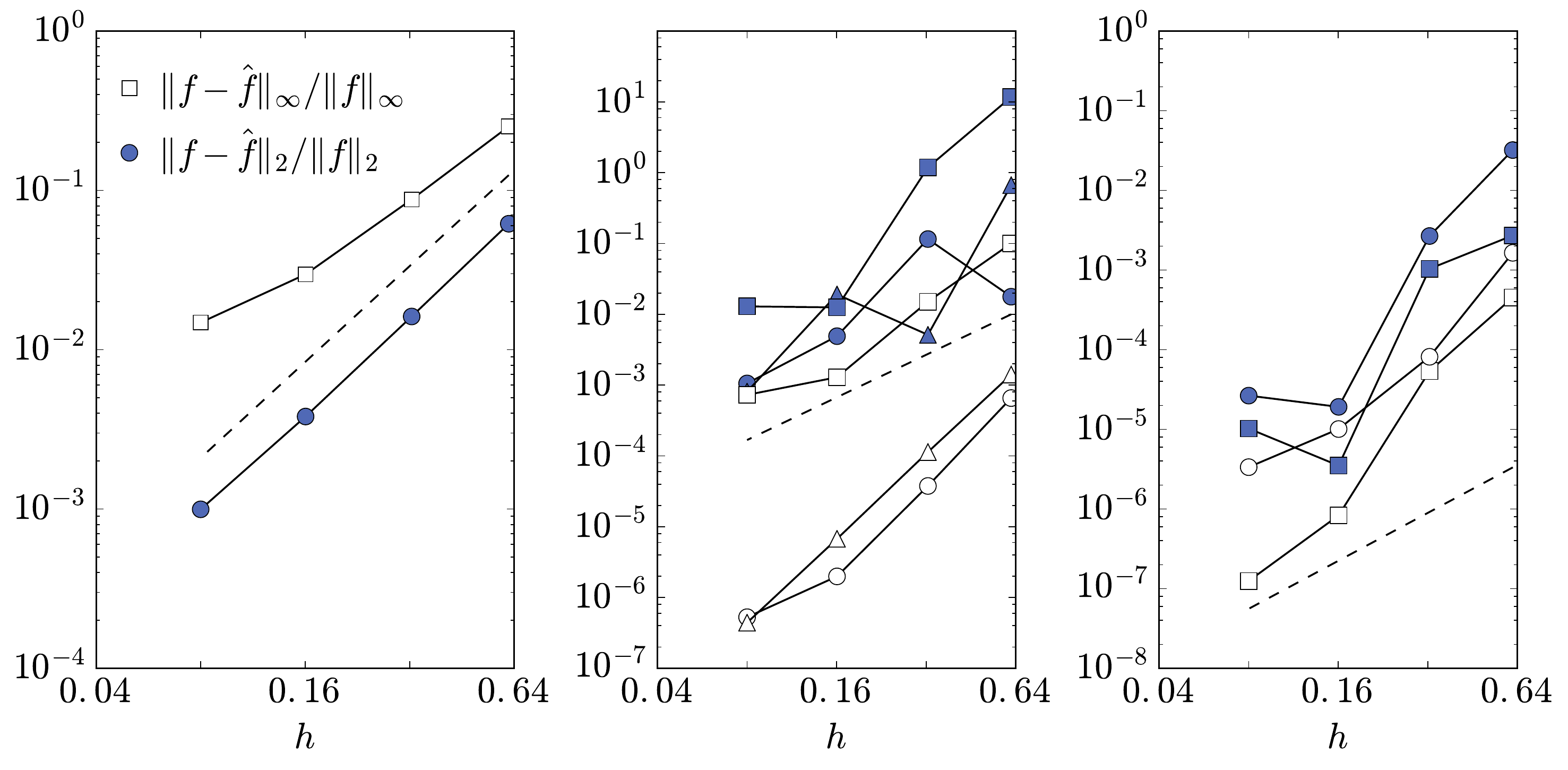}
\caption{Validation of the numerical method. Left panel: $L^{\infty }$ and
$L^{2}$ errors in the computed surface traction field on a sphere held immobile
in a background shear flow \tcb{in an unbounded domain} versus maximum triangle
edge length $h$ on a logarithmic scale. Center panel: \tcb{relative} error in
the net force (circles) and torque (squares) on a sphere translating parallel to
a wall, and torque (triangles) on a sphere rotating with rotation axis parallel
to the wall. The distance of the centroid to the wall is $H = \cosh (2)$ (far
from the wall; open symbols) or $H =\cosh (0.3)$ (near the wall; filled
symbols). Right panel: \tcb{relative} error in the force (squares) and torque
(circles) on a sphere held fixed in a background shear flow \tcb{near a wall}.
Open/filled symbols indicate the same distances to the wall as in the center
panel. Dashed lines scale as $h^{2}$ for reference, indicating second-order
accuracy of the numerical method.}
\label{fig:verification}\vspace{6pt}
\end{figure}

To demonstrate the accuracy of the free-space discrete system \reftext{\eqref{eq:discrete_free}}, we consider the case of a rigid sphere held
fixed in a background shear flow with a unit shear rate: $|\bm{u}(
\bm{x}) - x_{3}\bm{\hat{x}}_{1}|\to 0$ as $|\bm{x}|\to \infty $ and
$\bm{u}(\bm{x}) = \bm{0}$ for $|\bm{x}|=a$, for which an exact solution
is known \cite{kk91}. The analytical disturbance velocity field
may be written as the sum of a rotlet, a~stresslet, and the Laplacian
of a stresslet placed at the center of the sphere, resulting in
%
\begin{align}
\begin{split}
u_{i}(\bm{x})
&= \delta_{1i} x_{3} - \frac{a^{3}}{2}\frac{\epsilon
_{ijk}x_{k}}{r^{3}}\delta_{2j} - \left(  \frac{5a^{3}}{4} + \frac{a
^{5}}{8}\nabla^{2} \right) \frac{x_{i}x_{j}x_{k}}{r^{5}}s_{jk}
\\
&= \left( \frac{1}{2} - \frac{a^{3}}{2r^{3}}\right)  \epsilon_{i2k}x
_{k} + \left( \frac{1}{2} -\frac{a^{5}}{2r^{5}}\right) s_{ij}x_{j} +
\left( \frac{5a^{5}}{4r^{7}}-\frac{5a^{3}}{4r^{5}}\right) x_{i}x_{j}x
_{k}s_{jk},
\end{split}
\label{eq:exact_velocity_on_sphere_in_shear}
\end{align}
where $s_{jk} = \delta_{3j}\delta_{1k} + \delta_{1j}\delta_{3k}$. The
corresponding pressure is given by
%
\begin{align}
p(\bm{x})
&= 0 - \mu \left(  \frac{5a^{3}}{4} + \frac{a^{5}}{8}\nabla
^{2} \right)  \left( -2\frac{\delta_{jk}}{3r^{3}}+ 2\frac{x_{j}x_{k}}{r
^{5}} \right)  s_{jk} =\frac{-5\mu a^{3}}{2r^{5}}x_{j}x_{k}s_{jk},
\end{align}
and the tractions at the sphere surface are given by
%
\begin{align}
f_{i}(\bm{x}) =\left( -p(\bm{x})\delta_{im} + \mu \frac{\partial u_{i}}{
\partial x_{m}}+ \mu \frac{\partial u_{m}}{\partial x_{i}}\right)  \frac{x
_{m}}{a} = \frac{3\mu }{2a}\epsilon_{i2k}x_{k} + \frac{5\mu }{2a}s
_{ij}x_{j}
= \frac{\mu }{a}\left( 4\delta_{1i}x_{3} +
\delta_{3i}x_{1}\right) .
\label{eq:exact_traction_on_sphere_in_shear}
\end{align}
Although the body in this test is spherical, we wish to illustrate the
accuracy of the method in a more general setting and so we will not
exploit the symmetry in the problem, using instead a quadrature
obtained from a triangulated mesh as described above. The first panel
of \reftext{Fig.~\ref{fig:verification}} shows the relative $L^{2}$ and
$L^{\infty }$ errors between the exact traction field
\reftext{\eqref{eq:exact_traction_on_sphere_in_shear}}, with $\mu =1$
and $a = 1$, and the computed results $\hat{\bm{f}}$, obtained from
\reftext{\eqref{eq:discrete_free}}, plotted on a logarithmic scale
against the maximum mesh triangle edge length $h$. The decrease in the
$L^{2}$ error is proportional to $h^{2}$ as expected (a dashed line
with slope 2 is included for reference). The $L^{\infty }$ error is
also decreasing at nearly the same rate, but lags somewhat behind the
$L^{2}$ accuracy. This may be a consequence of the coarser meshes not
being subsets of the finer meshes; with a quadrature exploiting
axisymmetry we observe a more obvious connection between the $L^{2}$
and $L^{\infty }$ convergence rates (in fact we observed third-order
accuracy when using a symmetric mesh in this test).

As a second test of the code we consider a wall-bounded problem \reftext{\eqref{eq:discrete_wall}}, and compare the numerical results with exact
solutions for the net force and torque on a unit sphere in a viscous
fluid near a plane wall \cite{gcb67,Oneill1991,chaoui2003}. We
consider several resistance problems and for each problem we compute
solutions using several meshes, with the number of nodes ranging from
$54 \le N \le 3318$, and at two values of the height of the sphere
center above the wall, $H$. The absolute error in the net force
(circles) and torque (squares) on a sphere translating parallel to the
wall, and torque (triangles) on a sphere rotating with rotation axis
parallel to the wall are shown in the center panel of
\reftext{Fig.~\ref{fig:verification}}. The distance of the centroid to the wall
is $H = \cosh (2)$ (far from the wall; open symbols) and $H
=\cosh (0.3)$ (close to the wall; filled symbols). Again a dashed line
is included which scales as $h^{2}$ for reference, indicating
second-order accuracy. The right panel of \reftext{Fig.~\ref{fig:verification}}
shows the absolute error in the force (squares) and torque (circles) on
a sphere held fixed in a background shear flow, with open and filled
symbols indicating the same distances to the wall as in the center
panel. In all of the cases studied the errors decrease as the square of
the mesh diameter $h$ as expected, and the error naturally depends
strongly on the gap size between the particle and the wall. The number
of GMRES iterations required to achieve relative residuals of
$10^{-12}$ is modest compared to the system size, and in particular
decreases as the discretization is refined. On the other hand, we found
that it increases when the particle is close to a plane wall. See \reftext{Table~\ref{tbl:gmres}} for details.

\begin{table}
\caption{Number of GMRES iterations (without preconditioning or restarting) for
the discrete resistance problems whose convergence to exact solutions appears on
the right side of \reftext{Fig.~\ref{fig:verification}}. We solved three resistance
problems at each of the indicated centroid-wall distances $H$ and mesh vertex
counts $N$, all three using the same matrix but different right-hand sides. The
displayed iteration counts are for the translation problem; the counts for the
other right-hand sides differ by no more than two. The iteration count decreases
slightly when the grid is refined but increases with decreasing gap size.}
\label{tbl:gmres}
\centering 
\begin{tabular}{rccccc}
\hline
$H$ $\backslash$ $N$ &\;\;& 54 & 198 & 828 & 3318 \\
\hline
$\cosh(2)$              && 24 & 23   & 22  & 20 \\
$\cosh(0.3)$              && 53 & 63   & 64  & 52 \\
\hline
\end{tabular}
\end{table}

\section{Applications}\label{sec:applications}

\begin{figure}
 \begin{center}
\begin{tabular}{cc}
\includegraphics[width=0.4\linewidth]{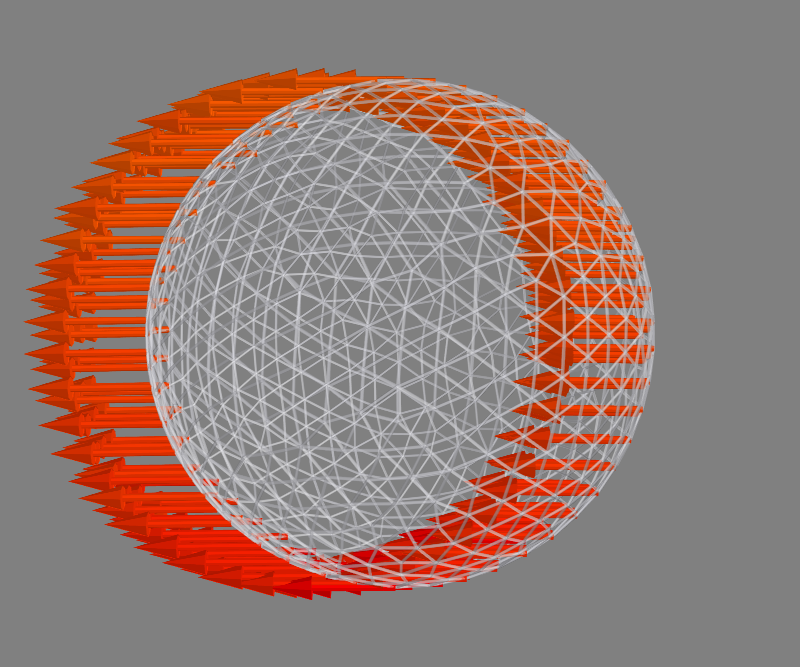} & \includegraphics[width=0.4\linewidth]{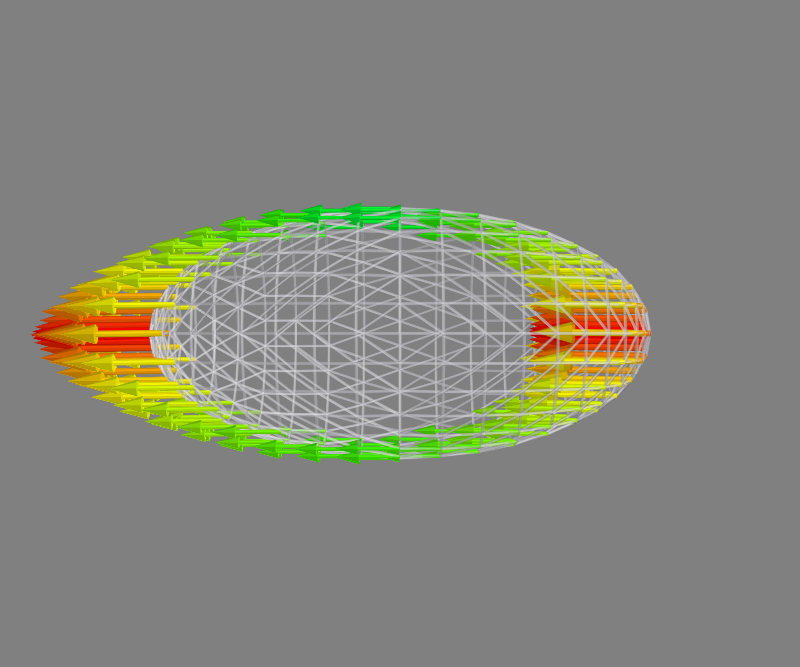}\\
\includegraphics[width=0.4\linewidth]{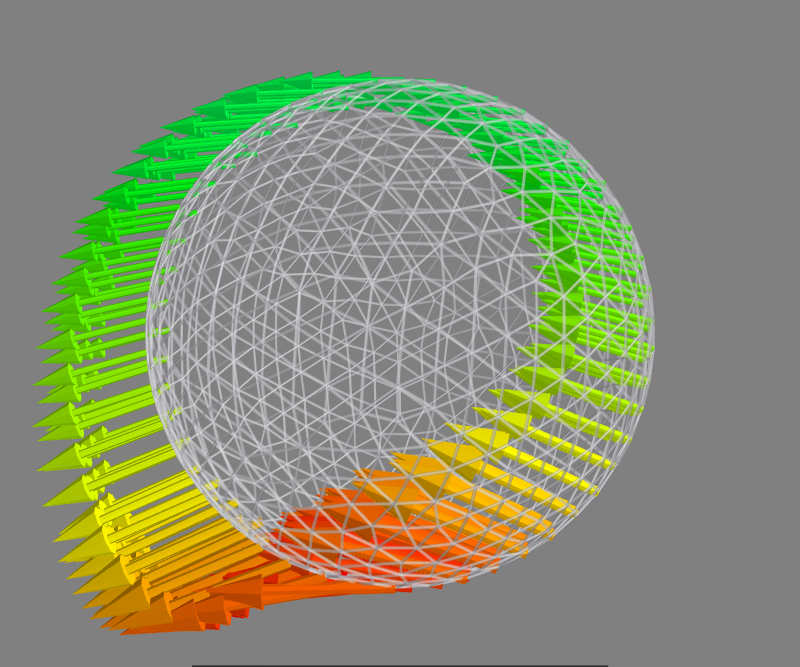} & \includegraphics[width=0.4\linewidth]{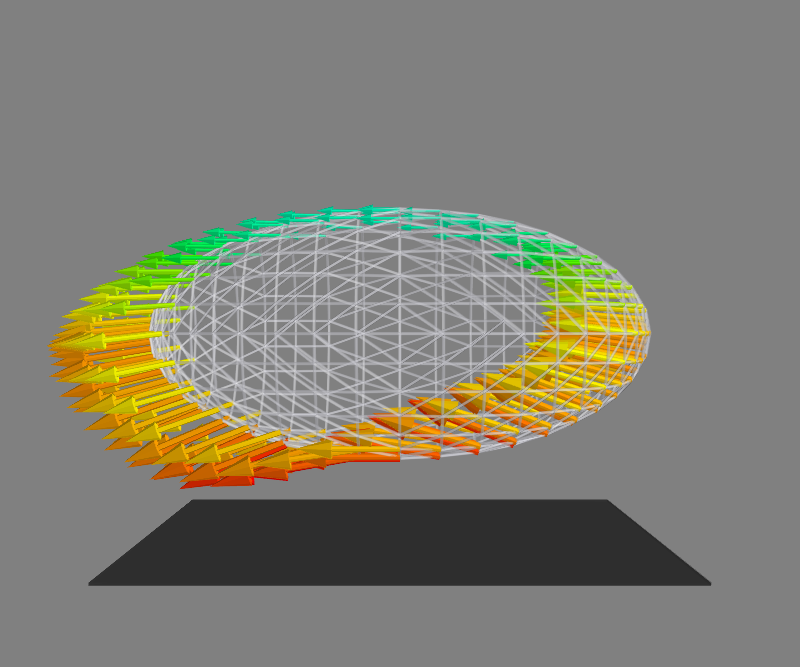}\\
\includegraphics[width=0.4\linewidth]{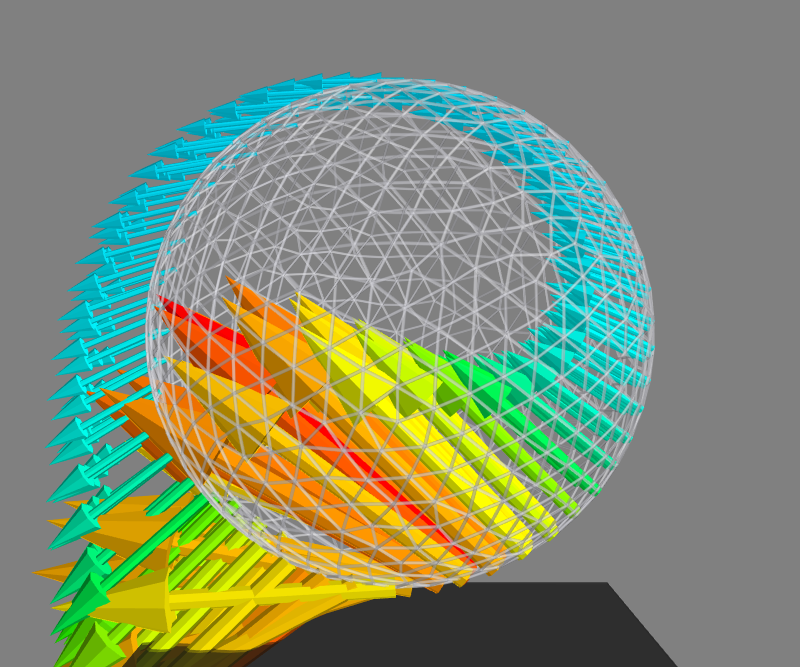} & \includegraphics[width=0.4\linewidth]{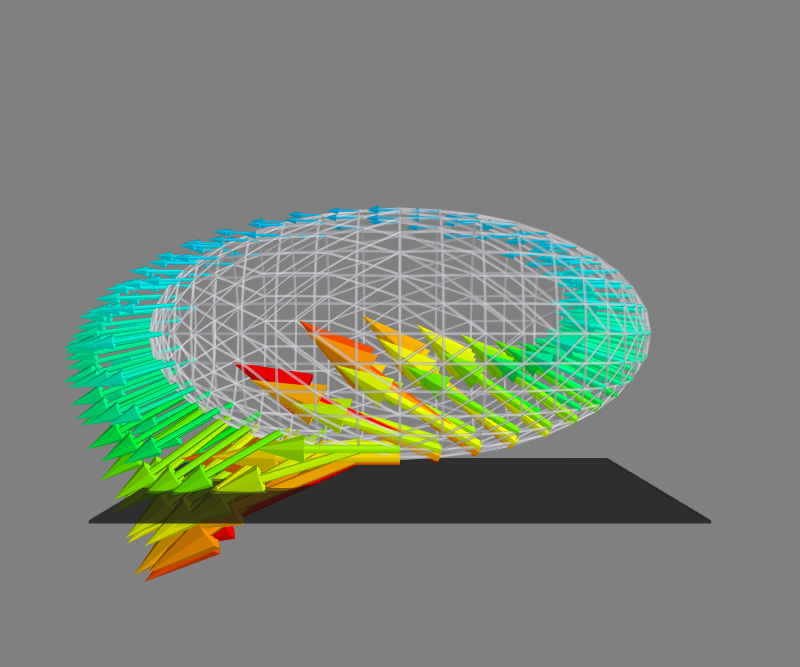}
\end{tabular}
\end{center}
\caption{Traction fields exerted by the fluid on torque-free particles \tcb{(left column: sphere; right column: prolate ellipsoid of aspect ratio 2)}
moving under a unit force to the right (parallel to the wall) at three different
distances to the wall. The wall is infinite in extent but represented visually
by a finite square. Tractions are plotted only at points near the $\{x_{2}=0\}$
plane for clarity. The traction fields are nearly up-down symmetric for
particles far from the wall, but symmetry is broken strongly by a nearby wall.
The resulting up-down asymmetry in the $x_{1}$-component of traction induces a
clockwise rotation on a torque-free body, whereas the front--back antisymmetry in
the $x_{3}$-component induces the opposite rotation. The net rotation due to the
combination of these two effects is clockwise for the case of a sphere and
counterclockwise for the prolate body, a~distinction which accounts for the
diverging trajectories of the two body shapes as previously studied
\cite{mitchell_spagnolie_2015}.}
\label{fig:lots_of_arrows}\vspace{6pt}
\end{figure}

\begin{figure}
 \begin{center}
\includegraphics[width=0.7\textwidth]{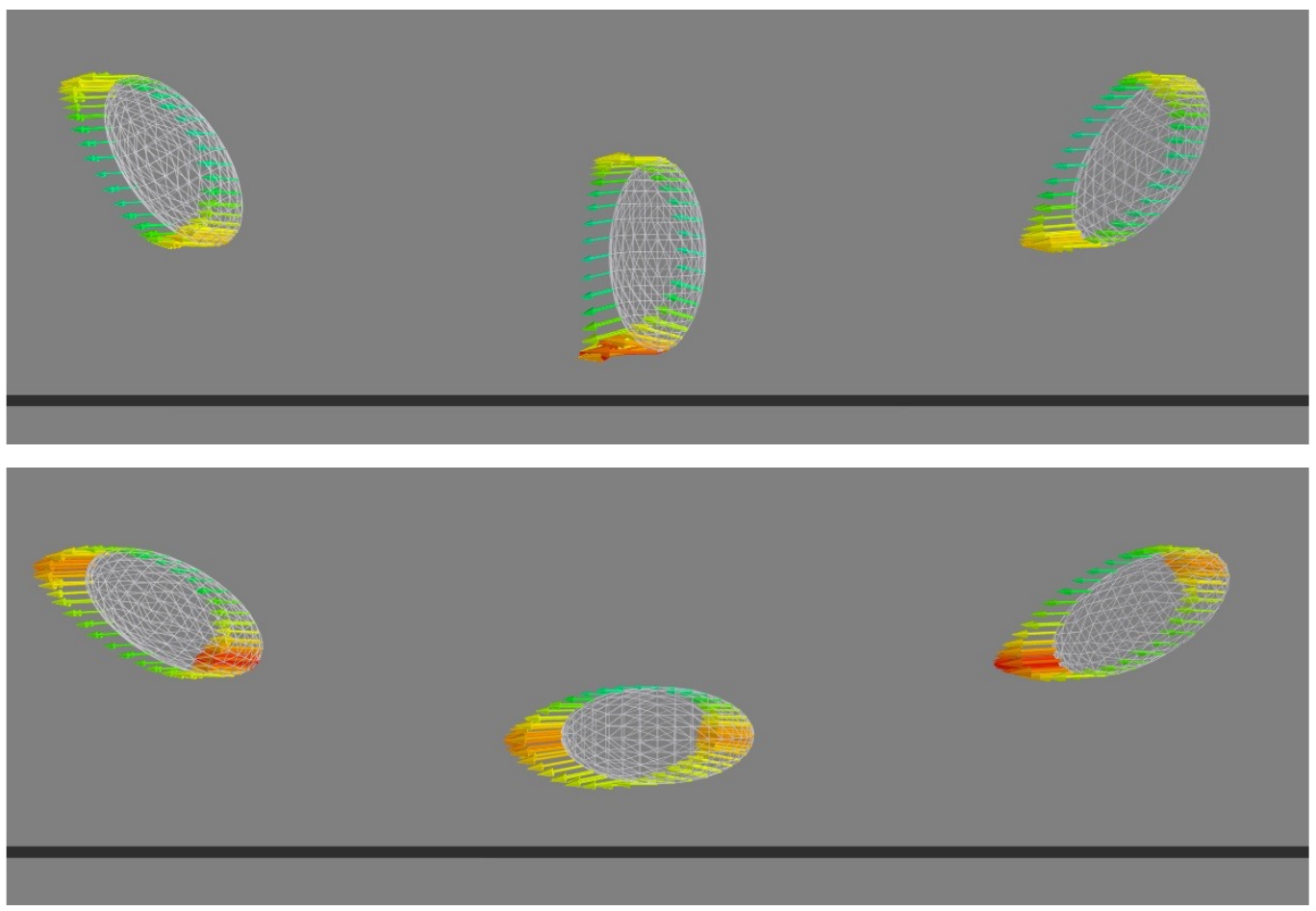} 
\end{center}
\caption{Top: a prolate ellipsoid moves above a no-slip wall with a constant
force to the right. Hydrodynamic traction vectors are shown with color and
length both in proportion to magnitude. Stresses are largest when the nose of
the body is closest to the wall, resulting in an overturning, or reversing
orbit. Bottom: The same body, with a smaller initial angle between its major
axis and the wall, undergoes instead a glancing orbit. When the body is closest
to the wall, the tractions on the body near the wall are still larger
than those on the side facing the bulk fluid, which would lead to a reversing rotation, but
there is a slight antisymmetry in the $x_{3}$-component of the traction at the
poles that instead results in an oppositely signed rotation.}
\label{fig:Reversing_vs_Glancing}\vspace{-2pt}
\end{figure}

\subsection{The role of particle eccentricity in \tcb{near-wall mobility}}\label{sec:arrows}

\tcb{The mobility of a rigid spheroid near a plane wall depends
strongly on the eccentricity of the particle. So too then does the
trajectory of such a body sedimenting under the influence of gravity,
and its diffusion under the action of Brownian fluctuations
\cite{lcw16}.} In a previous study we used far-field approximations to
produce analytical solutions for the \tcb{near-wall axisymmetric
particle mobility and resulting} body trajectory \tcb{in the context
of sedimentation} \cite{mitchell_spagnolie_2015}. Several
trajectory types are possible depending on the body shape and initial
position and orientation, including glancing, reversing, and periodic
tumbling. A~surprising result from that study was the finding that a
spherical body, which ``rolls'' down along a wall, is a very special
case. Indeed, while nearly spherical bodies periodically wobble towards
and away from the wall with a similar rolling-type rotation, a~more
eccentric prolate body with symmetry axis parallel to the wall can
instead rotate in the opposite direction; see
\reftext{Fig.~\ref{fig:lots_of_arrows}}. More concretely, the far-field theory
predicts this opposite rotation direction for a prolate body if the
height $H$ of its centroid above the wall satisfies $H^{2} > (2+e^{2})/(3e
^{2})$ where $e$ is the particle eccentricity
\cite{mitchell_spagnolie_2015}; such a body subsequently escapes from
the wall and recedes into the bulk fluid never to return. Generally,
escaping trajectories are classified as either glancing or reversing
(see \reftext{Fig.~\ref{fig:Reversing_vs_Glancing}}) depending on whether the
particle is parallel or perpendicular to the wall at the moment of
nearest approach. Which trajectory is selected depends on the particle
aspect ratio, and the initial orientation angle and height of the body
relative to the wall.

We now return to this problem to take a more detailed view of the local
hydrodynamic stresses which lead to body rotation near a wall. We place
a wall at $\{x_{3}=0\}$ and solve the mobility problem, imposing a unit
force in the $x_{1}$-direction on a rigid spherical particle or on a
prolate spheroidal particle whose symmetry axis is aligned with the
$x_{1}$-direction. \reftext{Fig.~\ref{fig:lots_of_arrows}} shows the computed
tractions on these bodies at three different distances from the wall.
For clarity, only the traction vectors at points near the $\{x_{2}=0
\}$ plane are shown. When the particle is infinitely far from the wall,
it is known that the traction on a sphere is uniform, and the traction
on a spheroid has a simple representation in terms of the surface normal
\cite{kim2015ellipsoidal}; in particular the traction fields on
both bodies at a large but finite distance from the wall are nearly
symmetric vertically and horizontally. Both symmetries break down at
smaller particle--wall gap sizes, and they break in such as way as to
introduce opposing torques, which in turn promote opposing rotations
since the particle is torque-free. The up-down asymmetry in the
$x_{1}$-component of the traction field corresponds to clockwise
rotation, whereas the front--back antisymmetry in the $x_{3}$-component
corresponds to counterclockwise rotation. For the sphere, the up-down
asymmetry is more powerful and the particle rotates in the clockwise
direction. The situation is reversed for the prolate body: the
front--back antisymmetry dominates because the locations of the strongest
$x_{3}$-tractions are better separated than in the spherical case,
resulting in a longer moment arm. A~similar competition between of
viscous stresses is observed throughout the typical glancing and
reversing trajectories shown in \reftext{Fig.~\ref{fig:Reversing_vs_Glancing}}.

\subsection{Viscous erosion}\label{sec:erosion}

As a second application we study the erosion of a body which is held
immobile in a uniform flow $\bm{u}^{\infty }(\bm{x}) = U \bm{\hat{x}}
_{1}$ or in a shear flow $\bm{u}^{\infty }(\bm{x}) = \dot{\gamma} x
_{3}\bm{\hat{x}}_{1}$. The surface is assumed to recede at a rate
proportional to the local shear stress,
%
\begin{equation}
\dot{\bm{x}} = -\alpha \big|(\bm{I} - \bm{n}\bm{n})\cdot \bm{f}\big|
\bm{n},
\label{eq:erosion_law}
\end{equation}
where $\bm{f}$ is the surface traction and $\alpha $ is a
proportionality constant. This ablation model was found to accurately
describe the erosion of a clay cylinder in an inertial flow
\cite{ristroph2012}. Assuming that the rate of particle erosion is small
when compared to the background flow, the shape dynamics may be studied
by a quasi-steady approximation. In other words, the no-slip velocity
boundary condition is applied on a rigid moving body to determine the
traction, which is used in \reftext{\eqref{eq:erosion_law}}.

We begin by simulating the erosion of an initially spherical body held
fixed in a uniform background flow in an infinite fluid. The problem is
made dimensionless by scaling lengths on the initial body radius
$a_{0}$, velocities on the background flow speed $U$, forces on
$\mu U a_{0}$, and time on a characteristic erosion timescale
$t_{v} = a_{0}^{2}/(\alpha \mu U)$. Henceforth all variables are assumed
to be dimensionless. By symmetry considerations the particle must remain
axisymmetric throughout the erosion process, and unlike in an inertial
flow it must also retain a fore-aft symmetry (seen by taking
$x_{1}\to -x_{1}$ and $u_{1}(\bm{x})\to -u_{1}(\bm{x})$ in \reftext{\eqref{eq:Stokes}}. The surface of the body at time $t$ is therefore
parameterized in a spherical coordinate system as described in
\S \ref{sec:numeric} with a radius function $r(\phi )\in C[0,\pi /2]$,
extended symmetrically for $\phi \in [\pi /2,\pi ]$ by $\tcb{r(
\phi )=r(\pi -\phi )}$. The function $r(\phi )$ is represented
numerically as a piecewise cubic spline on $[0,\pi /2]$ with 37 equally
spaced internal knots; the quadrature rule for the body surface is then
generated using 150 Gauss quadrature nodes on $[0,\pi ]$ and 18-point
trapezoidal integration in the azimuthal direction. The surface
tractions are computed from \reftext{\eqref{eq:discrete_free}} and the particle
shape is updated using \reftext{\eqref{eq:erosion_law}}, with timesteps chosen
adaptively so that unit normal vectors change by less than 0.001 in the
$L^{2}$ norm per step. After each timestep, all of the points on the
body surface are used to generate a new spline function representing the
updated profile $r(\phi )$.\looseness=-1

\begin{figure}[t]
 \begin{center}
 \begin{minipage}{0.25\linewidth}
  \includegraphics[width=\linewidth]{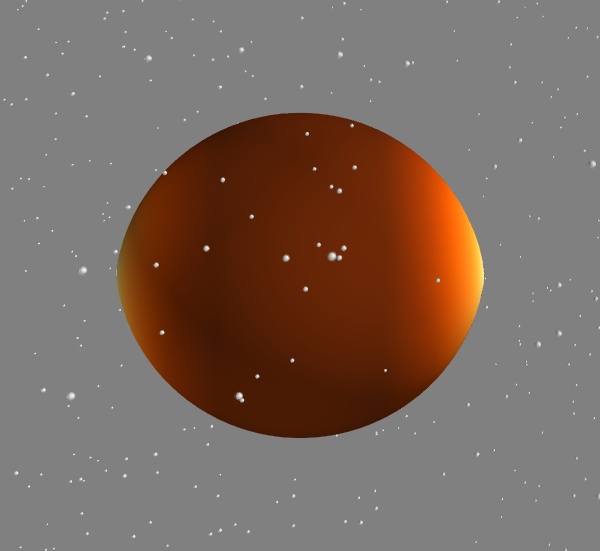}
 \end{minipage}  
 \begin{minipage}{0.25\linewidth}
  \includegraphics[width=\linewidth]{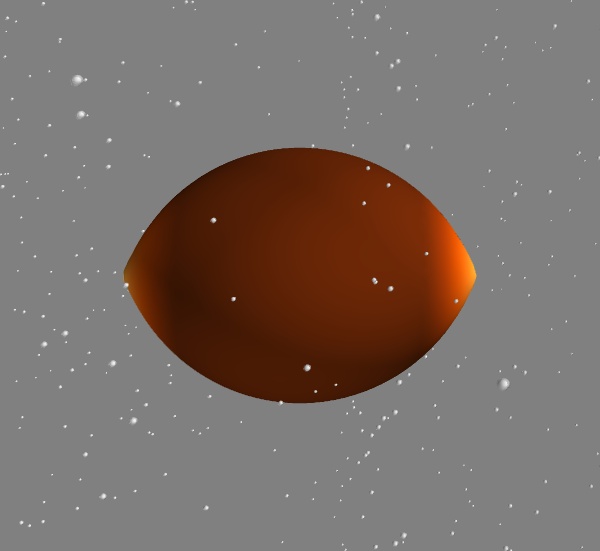}
 \end{minipage}  
 \begin{minipage}{0.25\linewidth}
  \includegraphics[width=\linewidth]{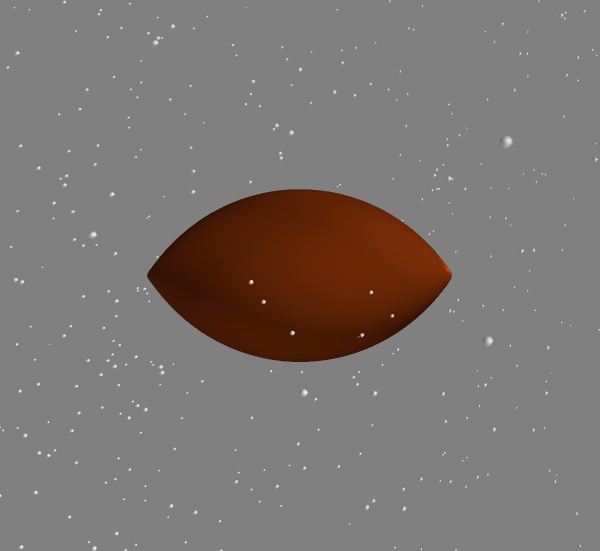}
 \end{minipage}  
\\
\vspace{.1in}
\includegraphics[width = 0.6\linewidth]{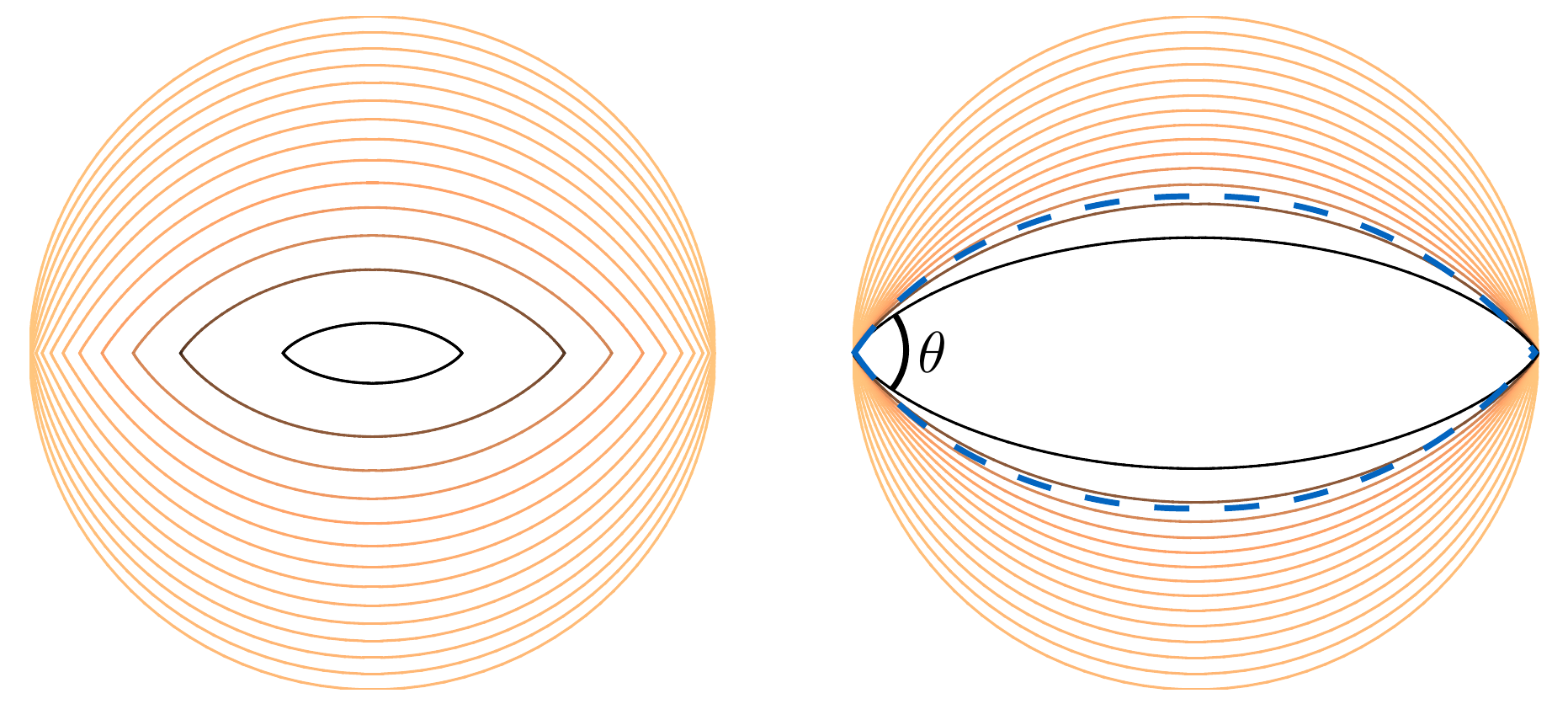}\\
\vspace{.1in}
\includegraphics[width = 0.9\linewidth]{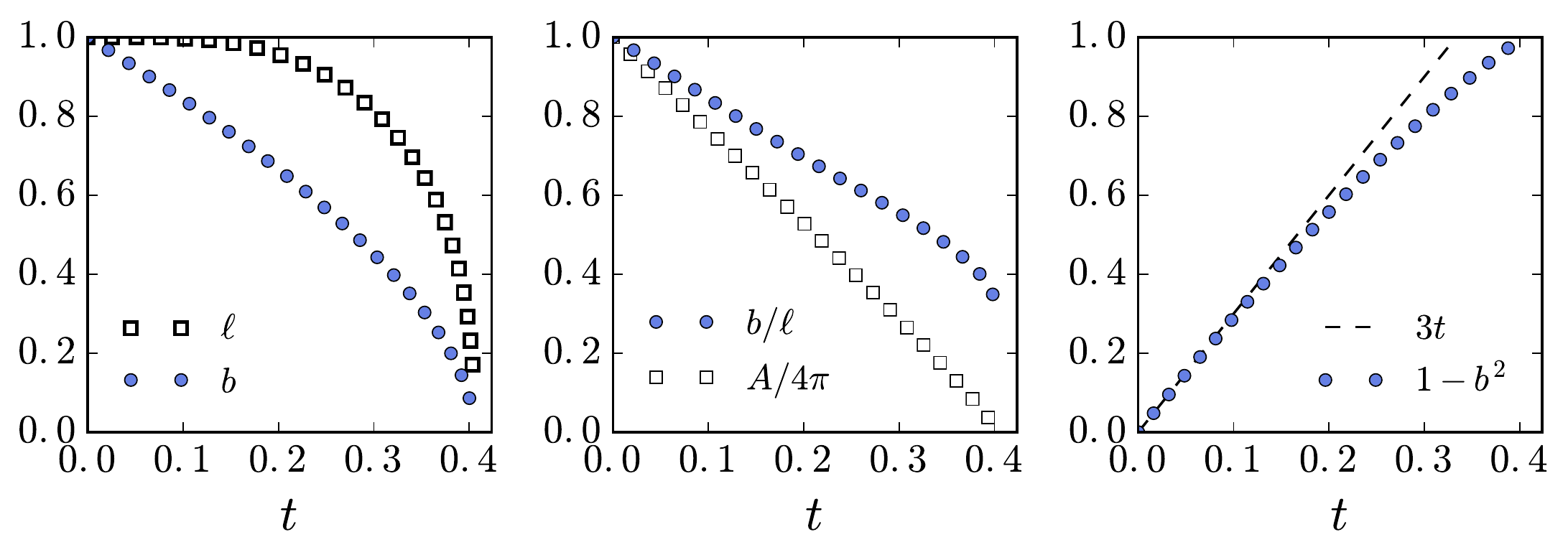}
 \end{center}
\caption{Erosion of an initially spherical particle at low Reynolds number in a
uniform horizontal background flow. Top row: the eroding particle at three
equally spaced times, with shear stress colored by shear stress magnitude
(darker color indicates a larger shear stress). Second row: cross-sectional
profiles taken through the $\{x_{2}=0\}$ plane at equally spaced times (left)
and scaled by the body half-length $\ell (t)$. The dashed line is Pironneau's
drag-minimizing profile \cite{Pironneau73,Bourot74}, which is not the limiting
body shape. 
Bottom row: The half-length $\ell (t)$ and half-thickness $b(t)$ as functions of dimensionless time (left), time-series of the body aspect ratio and surface area $A(t)$ (center), and comparison to theoretical prediction (right). }
\label{fig:profiles_uniform}\vspace{-3pt}
\end{figure}

\begin{figure}
 \begin{center}
\includegraphics[width = 0.8\linewidth]{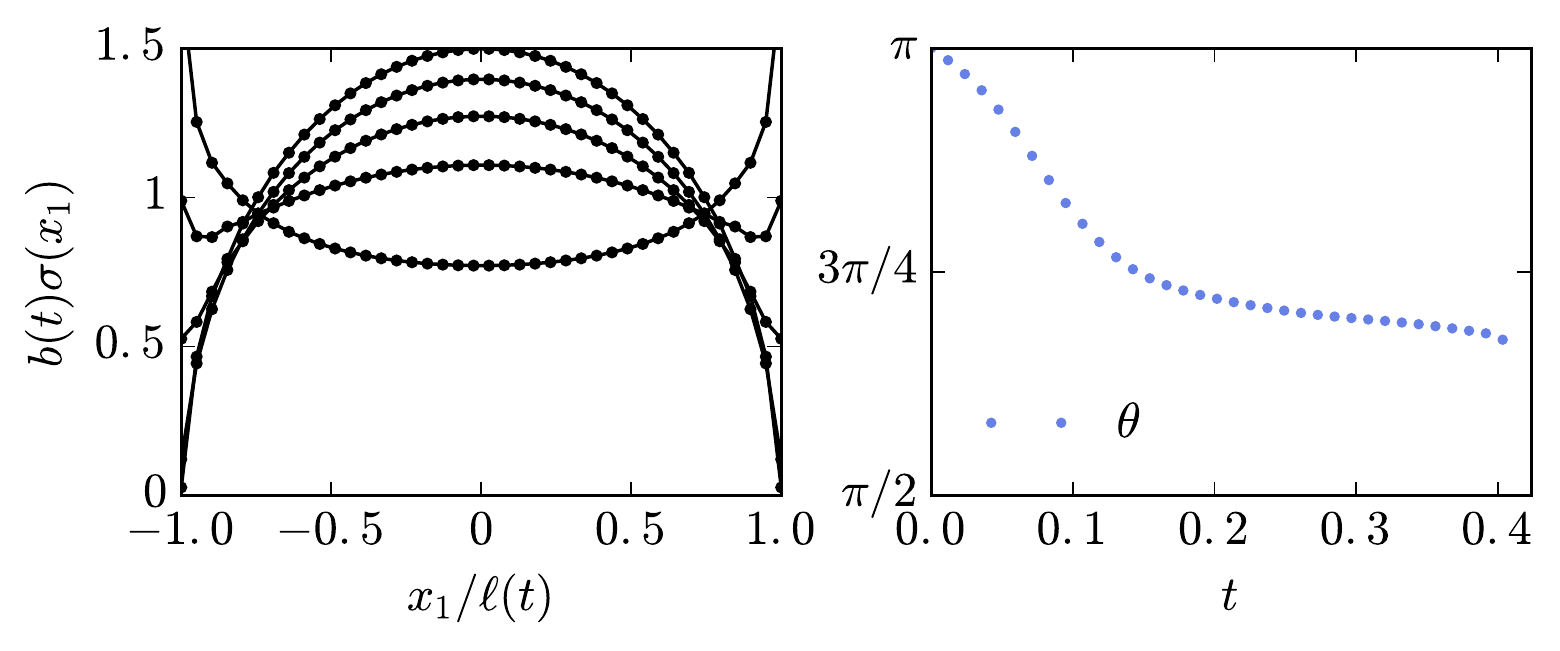}
\vspace{-.25in}
 \end{center}
\caption{Left: the thickness-scaled surface traction magnitude, $b(t)\sigma
(x)$, with $\sigma =|(\bm{I-nn})\cdot \bm{f}|$, at five equally-spaced moments
from $t=0$ to $t=0.4$ as a function of length-scaled horizontal position
$x_{1}/\ell (t)$, for the same simulation. \tcb{The tractions at the ends of
the particle diverge as the sharp features develop, in accordance with previous
analyses of Stokes flow past edges
\cite{mustakis1998microhydrodynamics}.} Right: the opening angle $\theta $ for
the eroding body in uniform flow depicted in \reftext{Fig.~\ref{fig:profiles_uniform}},
plotted against dimensionless time.}
\label{fig:thetas_tractions}
\end{figure}

The top row of \reftext{Fig.~\ref{fig:profiles_uniform}} shows the shape of the
eroding particle at three different times, with regions of increased
shear stress indicated by a darker color. As soon as the erosion begins
the body loses its smoothness at the leading and trailing stagnation
points, $\bm{x}=(\pm 1,0,0)$, instantly taking a locally conical
geometry there. Cross-sectional profiles of the body at equally spaced
moments in time are shown in the second row of
\reftext{Fig.~\ref{fig:profiles_uniform}}, in unscaled coordinates and then scaled
by the semiaxis length in the direction of flow, $\ell (t)$, to align
the horizontal axis lengths at all times. The first set of profiles
illustrate the accelerating reduction of length and width as the body
becomes smaller and as the aspect ratio decreases. The profiles are
again colored according to the magnitude of shear stress, and we note
in particular the gradual shift of the maximal shear stress from the
body midsection towards the conical endpoints. The shift of the maximal
shear stress from the midsection towards the nose and tip is
demonstrated more directly on the left in
\reftext{Fig.~\ref{fig:thetas_tractions}}, showing an initial profile which is
sinusoidal in the zenith angle $\phi $, which then becomes more evenly
distributed for intermediate times, and then finally inverts \tcb{so
that the tractions near the extremities become quite large, as is
appropriate given that Stokes flows past edges can produce
unbounded surface traction fields
\cite{mustakis1998microhydrodynamics}.} The rescaled profiles in
\reftext{Fig.~\ref{fig:profiles_uniform}} demonstrate that the aspect ratio of the
particle decreases monotonically throughout the evolution, a~result
which lies in contrast to the inertial study
\cite{ristroph2012} where the rescaled geometry converged to a
wedge-shaped region and then shrank in a self-similar fashion.

The half-length of the body, $\ell (t)$, and the half-thickness,
$b(t)$, are plotted as symbols on the bottom row of
\reftext{Fig.~\ref{fig:profiles_uniform}}. For small times $b(t)$ recedes at a
much faster rate than $\ell (t)$, but as the leading and trailing ends
become sharper the length reduction accelerates. Next the aspect ratio
$b(t) / \ell (t)$ is shown, indicating a steady monotonic decrease over
time with a very slight inflection at approximately $t = 0.2$, along
with the surface area $A(t)$, which shows a nearly linear decrease until
the particle vanishes. By the end of the simulation (which we halted
once $A(t)/A(0)<0.003$) the body reaches an aspect ratio of
approximately $0.3$. The simulations hint at a complete loss of volume
in finite time, and we do not expect the aspect ratio to reduce much
further before the body volume has vanished. Extrapolating from
$A(t)$ we predict a finite vanishing time of $t_{f}\approx 0.41$. The
opening angle $\theta $ is plotted as function of time on the right in
\reftext{Fig.~\ref{fig:thetas_tractions}}, where we observe a steady initial
decrease from the initial value of $\pi $ to a value just below
$3\pi /4$ before entering a period of much slower decay. The opening
angle does not settle to a fixed value in the time of simulation,
continuing to decrease very slowly, and reaching a value of
$\theta =0.67\pi $ by the end of the simulation. While the curves
showing the aspect ratio $b/\ell $ in \reftext{Fig.~\ref{fig:profiles_uniform}}
and $\theta (t)$ in \reftext{Fig.~\ref{fig:thetas_tractions}} suggest the
possibility of a rapid change before the body vanishes, a~more refined
study of this late stage indicated no such effect.

Many aspects of the evolution observed in the numerical simulation just
described can be understood better by analytical approximations. The
early loss of regularity is a consequence of the uniform dimensionless
traction $\bm{f} = (3/2)\bm{\hat{x}}_{1}$ as seen in the top left panel
of \reftext{Fig.~\ref{fig:lots_of_arrows}}. Indeed, the shear stress has magnitude
$\sigma (\phi ) = \big|(\bm{I} - \bm{n}\bm{n})\cdot \bm{f}\big|=3
\sin (\phi )/2$ (consistent with the shear stress plot at $t=0$ in
\reftext{Fig.~\ref{fig:thetas_tractions}}), so that by \reftext{\eqref{eq:erosion_law}} the
smooth surface develops a conical tip at $\phi =0$ via \tcb{$
\dot{\bm{x}}= -\sigma (\phi )\bm{\hat{n}}(\phi )\approx -3\phi /2
\bm{\hat{x}}_{1}$ for $\phi \approx 0$}. The opening angle is then given
by \tcb{a short exercise in trigonometry resulting in $\theta (t)
\sim \pi -3t$} for very small $t$.

Meanwhile, the erosion rate along the equator at the midsection of the
unit sphere is given by $\sigma (\pi /2)=3/2$. Consider the
approximation that this erosion rate applies uniformly on the body
surface. Then the particle remains spherical at all times with
dimensionless radius $a(t)$ (with $a(0)=1$), and the dimensionless
traction is given more generally by $\bm{f} = -3/(2 a(t))\bm{\hat{x}}
_{1}$. The radius would then decay as $a' = -3/(2a)$, so that
$a(t) = \sqrt{1 - 3t}$. This approximation indicates a finite time of
complete material loss, namely at a dimensionless time $t_{f} = 1/3$, a~smaller vanishing time than is observed in the simulations, but which
is quite a good estimate given the simplicity of the assumption. If
instead we consider the decay of the midsection alone, we might assume
for small times that the body length $\ell (t)$ remains roughly constant
but the thickness $b(t)$ decays according to the calculation above,
$b(t) = \sqrt{1 - 3t}$ or $1-b(t)^{2}=3t$. This approximation is
plotted in the rightmost panel of the bottom row of
\reftext{Fig.~\ref{fig:profiles_uniform}} as a dashed line, showing a close
agreement with the simulations over a significant length of
dimensionless time. The nearly linear decrease in the surface area
$A(t)$ on the bottom right of \reftext{Fig.~\ref{fig:profiles_uniform}} is also
consistent with this scaling. 
The predicted linear scaling in the surface area, proportional to $(t_f - t)$ for $|t_f-t| \ll 1$, sits in contrast to the power law scaling in a high Reynolds number flow \cite{ristroph2012,moore2013self}, though a more detailed mathematical investigation of the vanishing regime is still needed.  

It would seem plausible that the shape of the eroding sphere should
\tcb{tend} towards a limiting, self-similar profile as was found for
the eroding body in an inertial flow \cite{ristroph2012}, wherein
the shear stress is uniform so that the shape is maintained while the
volume shrinks. In particular, the drag-minimizing profile of a body
with fixed volume in a uniform Stokes flow was shown by Pironneau to be
that which everywhere has a constant vorticity (and hence shear stress)
\cite{Pironneau73}, resulting in a body with aspect ratio of
approximately $0.477:1$ and conical endpoints with an opening angle of
$2\pi /3=120^{\circ }$ \cite{Bourot74}. This profile is included
for reference as a dashed line in \reftext{Fig.~\ref{fig:profiles_uniform}}. Why
doesn't the body maintain this shape as it vanishes? To answer this we
now show that the only profiles preserved under a constant erosion in
the normal direction are those of a sphere and a cone.

Parameterizing the surface of an axisymmetric body as above by a
function $r(\phi )$ describing the distance from the particle center as
a function of the polar angle, we note that the condition $\dot{r}(
\phi ) \propto r(\phi )$ describes self-similar evolution. If the shear
stress is given by $\alpha \chi (\phi )$, then \reftext{\eqref{eq:erosion_law}}
becomes $\dot{\bm{x}}(\phi ) = -\alpha \chi (\phi )\bm{n}(\phi )$ and
we find the following condition for self-similar decay:
\begin{eqnarray*}
-\chi (\phi )\sqrt{1+\left( r'(\phi )/r(\phi )\right) ^{2}} = \beta r(
\phi ),
\end{eqnarray*}
where the decay parameter $\alpha $ has been folded into the constant
of proportionality $\beta $. Equivalently, a~profile $r(\phi )$ which
recedes in the normal direction at a rate proportional to $\chi (
\phi )$ remains self-similar if and only if
\begin{eqnarray*}
\frac{d}{d\phi }\left( -\chi (\phi )\sqrt{r(\phi )^{-2}+r'(\phi )^{2}
r(\phi )^{-4}}\right)  = 0,
\end{eqnarray*}
or
%
\begin{equation}
\chi '(\phi ) r(\phi )\Big(r(\phi )^{2} + r'(\phi )^{2}\Big) = \chi (
\phi )r'(\phi )\Big(r(\phi )^{2} - r'(\phi )r(\phi )+ 2r'(\phi )^{2}
\Big).
\end{equation}
If the erosion rate is constant over the whole body, as it is for the
drag-minimizing shape of Pironneau, then $\chi '(\phi )$ vanishes and
we have
%
\begin{equation}
0 = r'(\phi )\left[ -r(\phi )^{2} - 2r'(\phi )^{2} + r(\phi )r''(
\phi )\right] .
\end{equation}
The profile corresponding to $r'(\phi )=0$ is the sphere and the other
factor leads to the second-order ODE\nopagebreak
\begin{eqnarray*}
0=-r(\phi )^{2}-2r'(\phi )^{2} + r(\phi )r''(\phi ),
\end{eqnarray*}
which has a two-parameter family of solutions given by $r(\phi ) = c
_{1}c_{2}\left[ c_{2}\cos (\phi )+c_{1}\sin (\phi )\right] ^{-1}$, which
is the equation in polar coordinates of a straight line with intercepts
$(0,c_{2})$ and $(c_{1},0)$. This shows that if there is a limiting
profile under erosion induced by uniform background flow then that
limiting profile does not have constant shear stress, and in particular,
that the profile of Pironneau cannot be the limiting shape of a particle
eroding under the ablation law \reftext{\eqref{eq:erosion_law}}. Generally, a~uniform interface velocity on a body with variable curvature does not
result in a self-similar shape reduction, a~point which was also made
in the context of erosion at high Reynolds numbers
\cite{moore2013self}.

The question remains whether there is a limiting profile of smaller
aspect ratio than those achieved by the initially spherical particle in
the time before dissolution. We also simulated the erosion of initially
prolate bodies in a uniform background flow, but found that the aspect
ratio always decreases monotonically even when the initial ratio is as
small as 1:5. On this basis we conjecture that no profile eroding in a
uniform flow according to \reftext{\eqref{eq:erosion_law}} can shrink in a
self-similar fashion.

\subsubsection{Erosion in background shear flows and near surfaces}\label{sec:shear_erosion}
We now turn to the case of an immobile and initially spherical particle
eroding under the influence of a background shear flow. Three
configurations are considered: when the initial particle center lies in
the plane of zero background flow in an infinite fluid, when it lies
$1.5$ radii above the plane of zero background flow, and finally when
it lies $1.5$ radii above a no-slip planar boundary. For the geometry
discretization we begin with a uniform triangulation from \texttt{distmesh}
which we then evolve without retriangulation according to \reftext{\eqref{eq:erosion_law}}. At each step, the normal vectors and quadrature
weights are recomputed as described in \S \ref{sec:numeric}.
Nondimensionalization is similar to that applied in the uniform flow, but with velocities scaled upon $\dot{\gamma}a_0$, where $a_0$ is the initial particle size and $\dot{\gamma}$ is the shear rate.

\begin{figure}
\begin{center}
\begin{minipage}{0.75\linewidth}
\begin{tabular}{ccc}
  \includegraphics[width=0.3\linewidth]{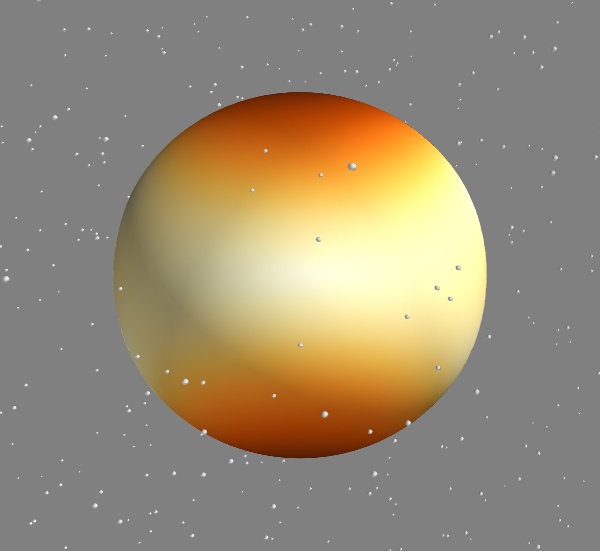} &
  \includegraphics[width=0.3\linewidth]{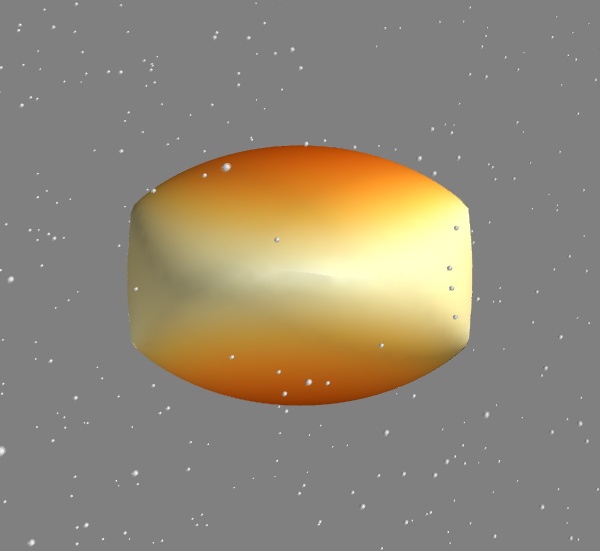}&
  \includegraphics[width=0.3\linewidth]{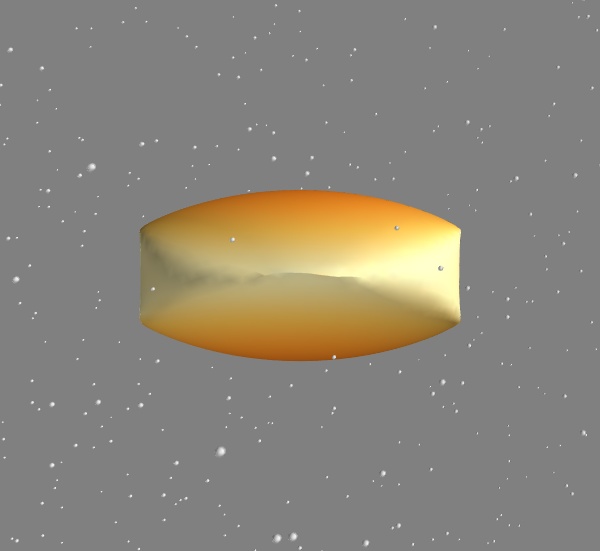}\\
  \includegraphics[width=0.3\linewidth]{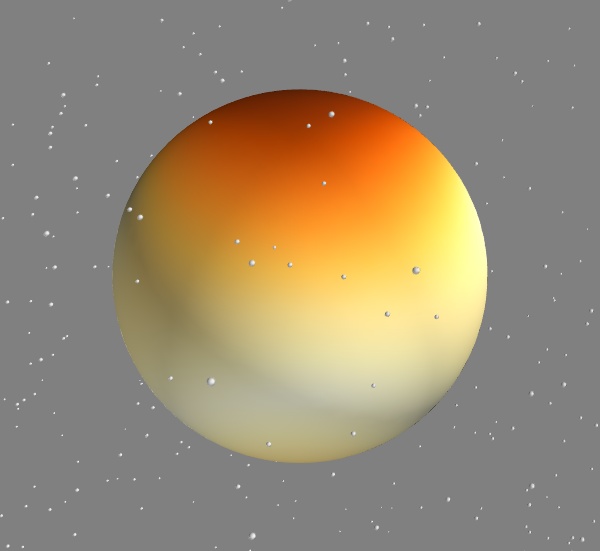} &
  \includegraphics[width=0.3\linewidth]{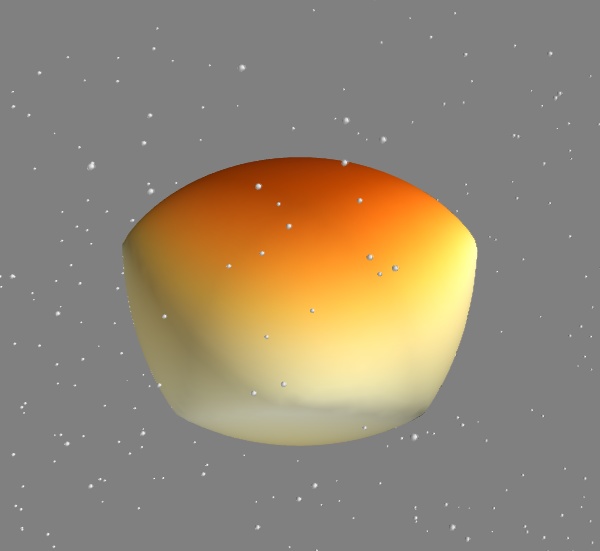}&
  \includegraphics[width=0.3\linewidth]{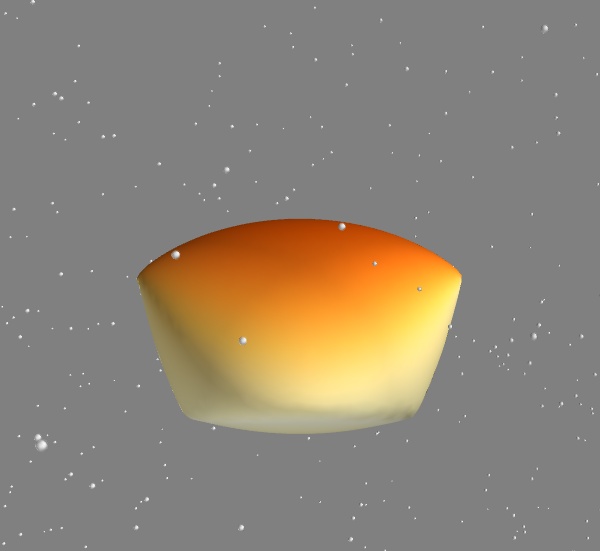}\\
  \includegraphics[width=0.3\linewidth]{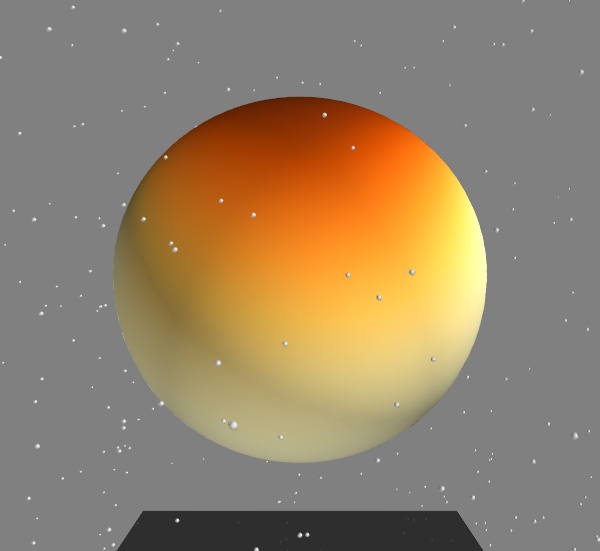}&
  \includegraphics[width=0.3\linewidth]{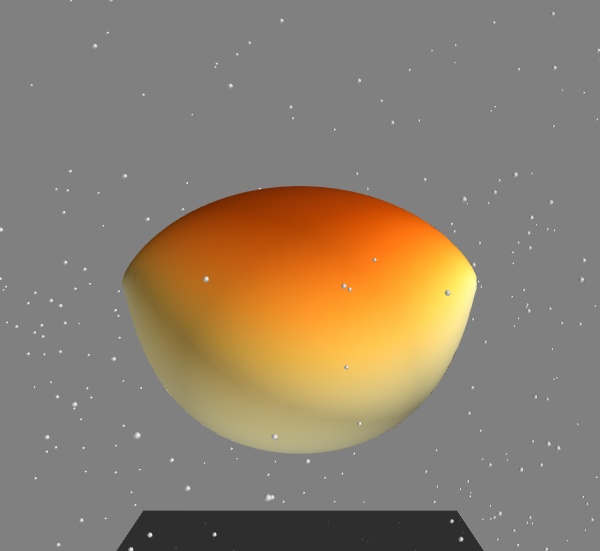}&
  \includegraphics[width=0.3\linewidth]{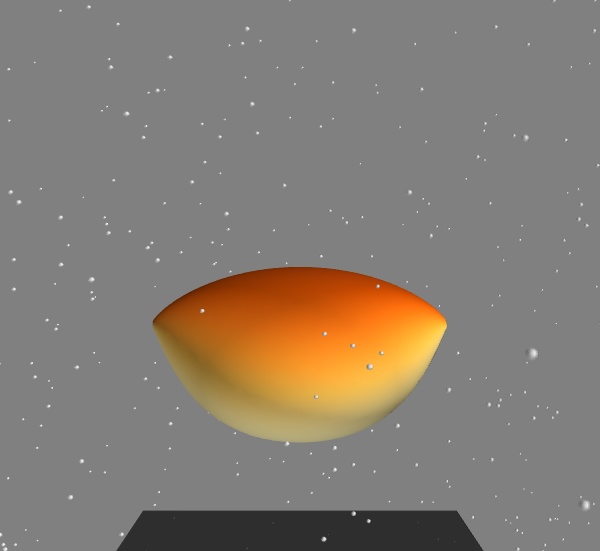} 
\end{tabular}
\end{minipage}~\begin{minipage}{0.25\linewidth}
\includegraphics[width=\linewidth]{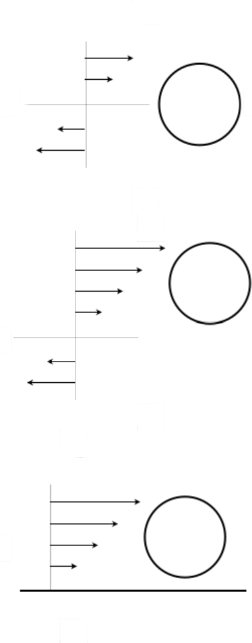}
\end{minipage}
 \end{center}
\caption{Erosion of initially spherical particles at low Reynolds number while
held immobile in a background shear flow. Top row: the fluid is infinite and the
plane of zero shear flow meets the particle center. Second row: the fluid is
infinite and the plane of zero shear lies $1.5$ radii below the initial particle
center. Bottom row: the fluid is bounded by a no-slip wall $1.5$ radii below the
initial particle center.}
\label{fig:movie_with_tracers}\vspace{-3pt}
\end{figure}

\begin{figure}
 \begin{center}
  \begin{tabular}{c c c}
\includegraphics[width=0.3\linewidth]{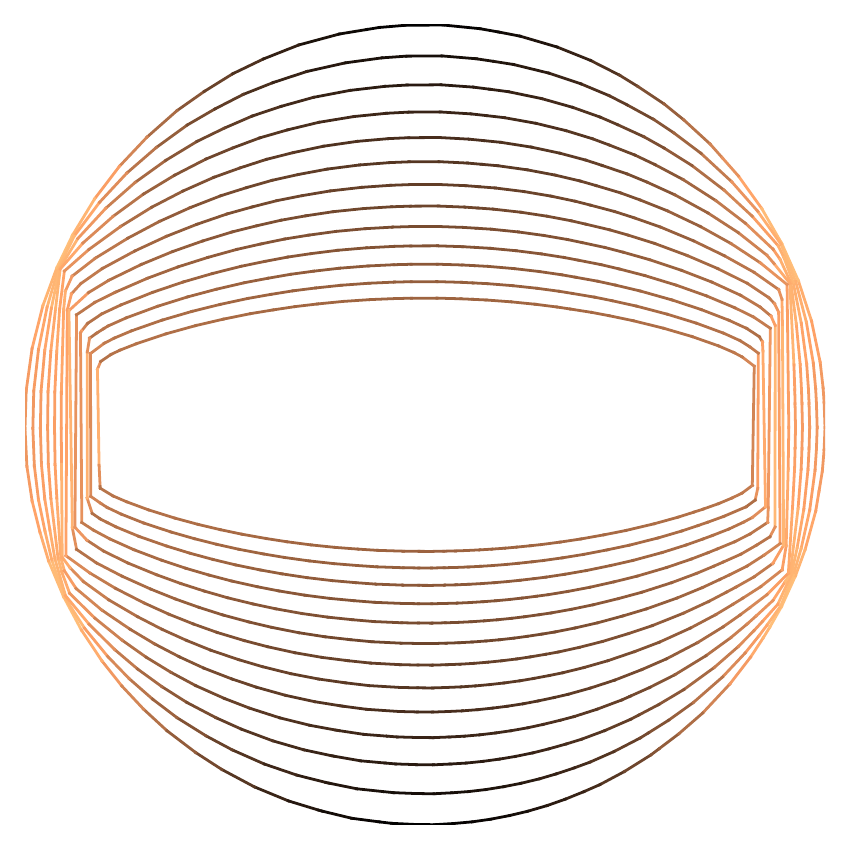}&
   \includegraphics[width=0.3\linewidth]{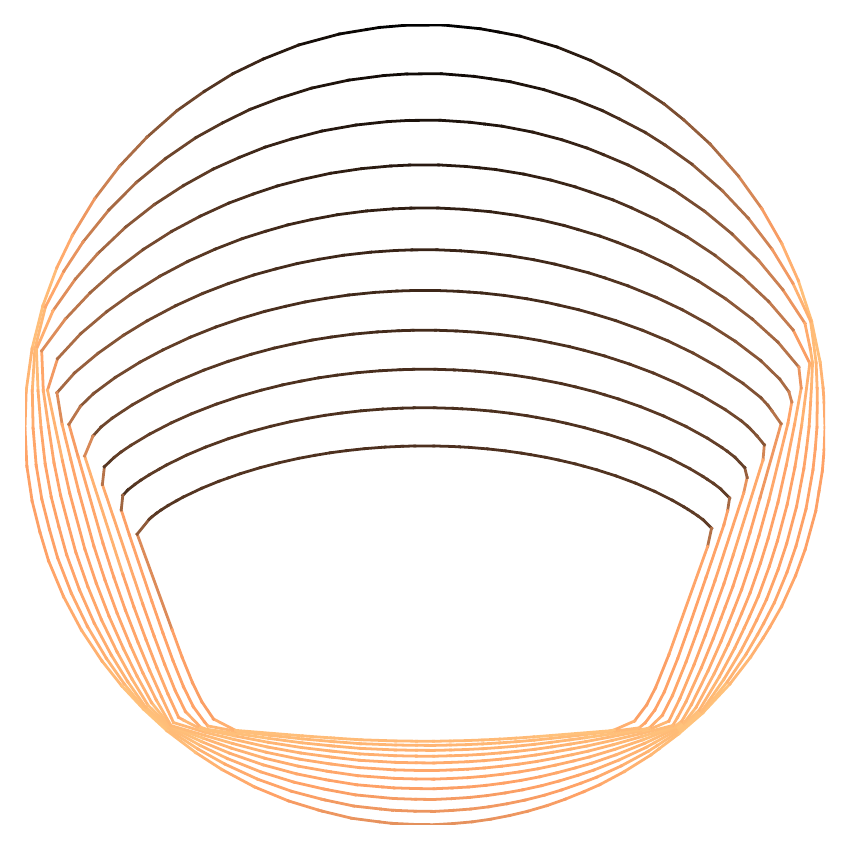}&   
   \includegraphics[width=0.3\linewidth]{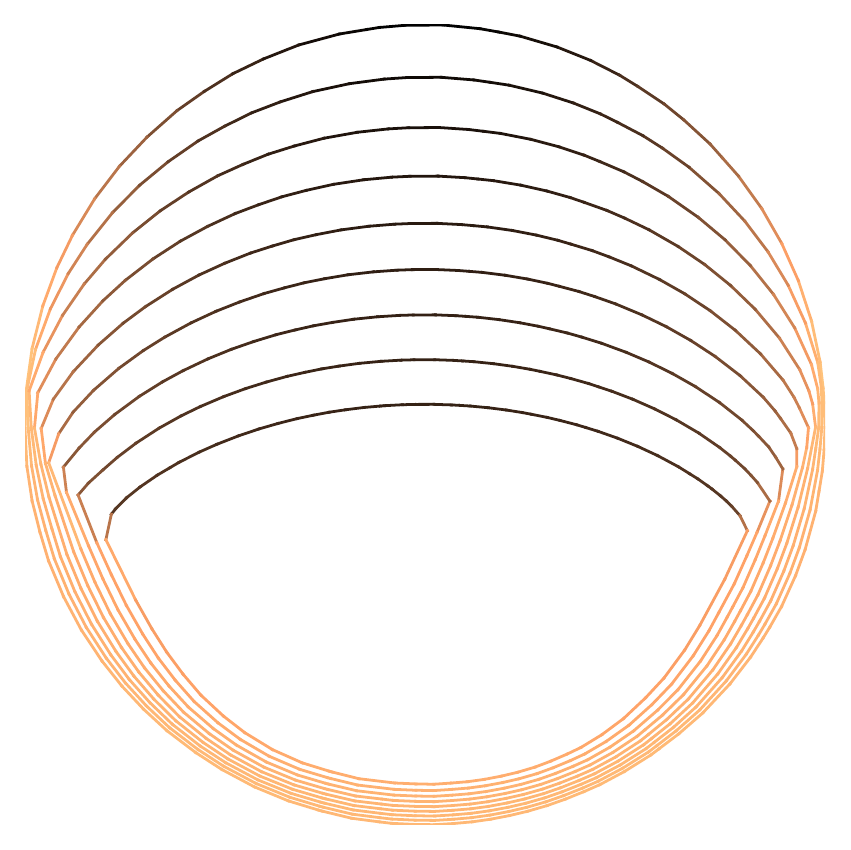}\\  
   \includegraphics[width=0.3\linewidth]{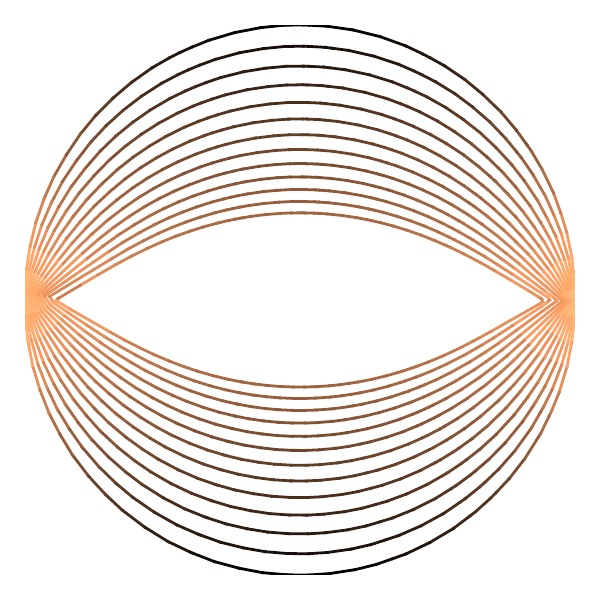}&
   \includegraphics[width=0.3\linewidth]{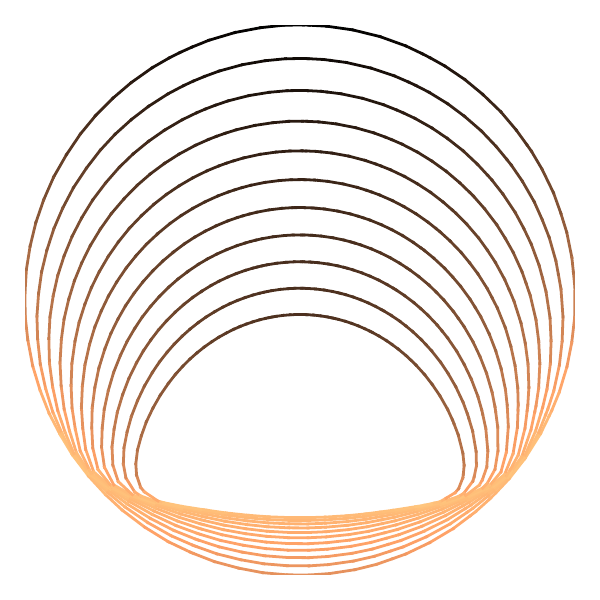}& 
   \includegraphics[width=0.3\linewidth]{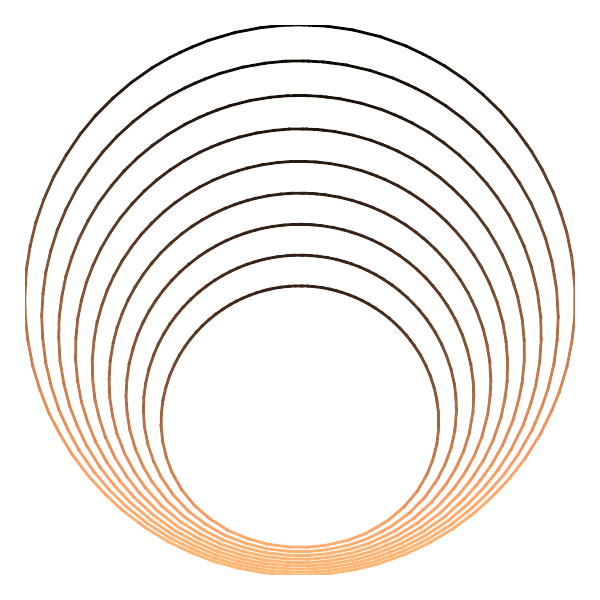}\\  
  $(a)$ & $(b)$ & $(c)$   \\
   \end{tabular}
 \end{center}
 \caption{Cross-sectional profiles of eroding bodies in a background shear flow
as in \reftext{Fig.~\ref{fig:movie_with_tracers}}, equally
spaced in time. 
The background flow is in the $x_{1}$-direction and varies
linearly in the $x_{3}$-direction; the $x_{2}$-direction is neutral. 
Shown are
views from the neutral direction (top) and along the flow direction (bottom) for
an initially spherical body whose centroid is located (a) in the plane of zero shear flow in an unbounded fluid; $(b)$ $1.5$ radii above the plane of zero shear flow in an unbounded
fluid; and $(c)$ $1.5$ radii above a no-slip wall. Darker
color indicates higher shear stress; the color scaling changes across rows but
not within columns. Counting all of the sharp regions in each column, we find
six, four, and two for $(a)$, $(b)$, and $(c)$, respectively.}
\label{fig:outlines}\vspace{-3pt}
\end{figure}

In the first problem the particle is placed at the origin and we let
$\bm{u}^{\infty }(\bm{x}) = x_{3}\bm{\hat{x}}_{1}$. The center of the
particle then lies in the plane of zero background flow, resulting in
top-down symmetry in addition to the ever-present fore-aft symmetry.
Axisymmetry, however, is now broken by the background flow. The
resulting evolution is depicted in the first row of
\reftext{Fig.~\ref{fig:movie_with_tracers}} and a series of cross sections viewed
along the $x_{1}$- and $x_{2}$-directions are provided in
\reftext{Fig.~\ref{fig:outlines}}(a). Here again darker colors indicate locations
of increased shear stresses. The surface regions facing the direction
of flow experience relatively small viscous stresses at early times and
the particle flattens due to the larger stresses near the extremal
points in the direction of shear. In fact there is a recirculation
region near the front and back of the particle relative to the flow
direction resulting in a non-convex profile there. This slowly eroding
non-convex region meets the more rapidly eroding regions on the top and
bottom in a pair of sharp features, best viewed from the $x_{2}$-direction in
the top image of \reftext{Fig.~\ref{fig:outlines}}(a). Meanwhile, along the
neutral axis, the surface points with extremal $x_{2}$-values develop
into cusps, as can be seen in the profiles viewed from the
$x_{1}$-direction in \reftext{Fig.~\ref{fig:outlines}}. In the later stages of the
evolution there are six singular features, four in the $\{x_{2}=0\}$ plane and two
in the $\{x_{1} = 0\}$ plane.

\begin{figure}
\begin{center}
\begin{tabular}{ccc}
 \includegraphics[width = 0.4\linewidth]{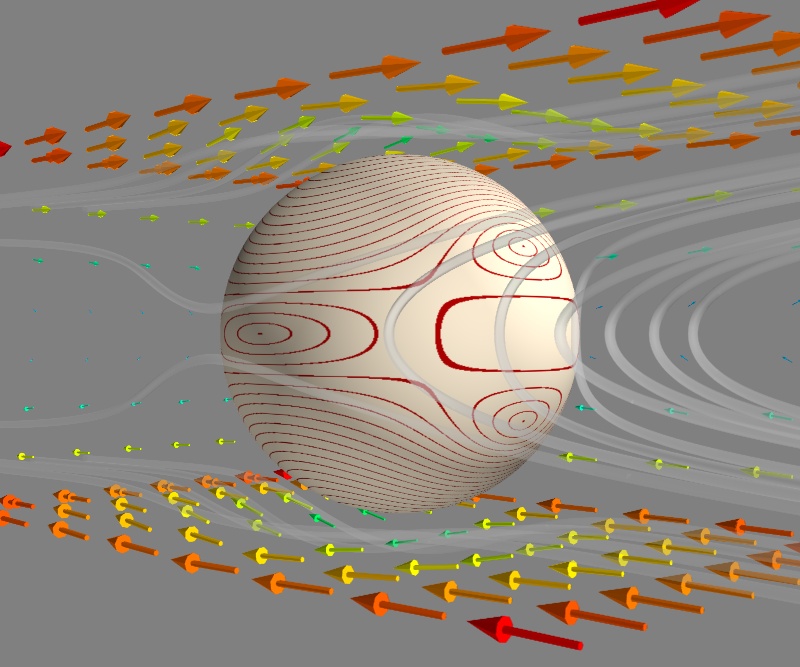} &
\; &
 \includegraphics[width = 0.4\linewidth]{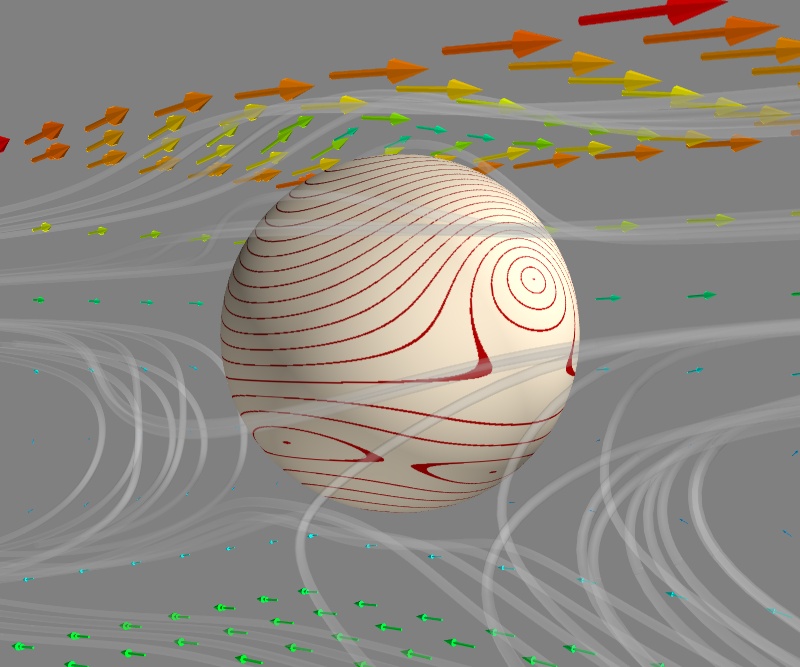}\\
 (a) & & (b)
 \end{tabular}
 \end{center}
\caption{Exact flow fields and surface tractions for a sphere held fixed (a)
with the plane of zero background flow through the sphere center, and (b) with
the plane of zero background flow located $1.5$ radii below the sphere center
(as in the first two rows of \reftext{Fig.~\ref{fig:movie_with_tracers}}). Contours of the
shear stress magnitude, $|(\bm{I} - \bm{n}\bm{n})\cdot \bm{f}|$ are shown
on the spherical surface. In both cases we see regions of recirculating flow
(lower in $(b)$ than in $(a)$), and in both cases the locations of zero shear
stress are stagnation points for the flow, resulting in an impending loss of
surface smoothness.}
\label{fig:traction_contours}
\end{figure}

Contours of the exact shear stress on the initially spherical body,
along with streamlines of the exact fluid velocity at $t=0$ (see
\reftext{\eqref{eq:exact_velocity_on_sphere_in_shear}}) are shown in
\reftext{Fig.~\ref{fig:traction_contours}}(a). This initial flow field
has regions of recirculation between the stagnation points $(\pm
2/\sqrt{5}, 0, \pm 1/\sqrt{5})$, obtained from setting the tangential
component of the traction field
\reftext{\eqref{eq:exact_traction_on_sphere_in_shear}} to zero. This
offers another perspective on why the surface recedes both in the
recirculation region and, more rapidly, on the high-shear region at the
poles, and it is highly suggestive of the nature of the eroded shape
and locations of regularity loss to come.

As a second problem we again begin with a spherical particle centered
at the origin, but now we take background velocity $\bm{u}^{\infty }(
\bm{x}) = (x_{3}+\kappa )\bm{\hat{x}}_{1}$ so that the sphere center
lies a distance $\kappa $ above the plane of zero background flow. For
$\kappa >1$ the background flow is everywhere in the positive
$x_{1}$-direction over the body surface, and the flow is stronger at the
top of the particle. The particle evolution is shown in the second row
of \reftext{Fig.~\ref{fig:movie_with_tracers}} and a series of outlines are
plotted in \reftext{Fig.~\ref{fig:outlines}}(b). The most dramatic change compared
to the more symmetric configuration considered above is the absence of
sharp features in the profile viewed along the $x_{1}$-direction; i.e. the
second column of \reftext{Fig.~\ref{fig:outlines}} has fewer sharp features than
the first. To better understand the shape evolution we again look to the
exact velocity and traction fields at $t=0$. The exact velocity and
traction fields at $t=0$ are found by adding to \reftext{\eqref{eq:exact_velocity_on_sphere_in_shear}} the flow due to a Stokeslet
and source dipole, resulting in the fluid velocity
%
\begin{eqnarray}
u_{i}(\bm{x}) &=& \left( \frac{1}{2} - \frac{a^{3}}{2r^{3}}\right)
\epsilon_{i2k}x_{k} + \left( \frac{1}{2} -\frac{a^{5}}{2r^{5}}\right) s
_{ij}x_{j} + \left( \frac{5a^{5}}{4r^{7}}-\frac{5a^{3}}{4r^{5}}\right) x
_{i}x_{j}x_{k}s_{jk}
\nonumber
\\
&&{}
+\kappa \delta_{1i}\left( 1-
\frac{3}{4}\left( \frac{a}{r} + \frac{a^{3}}{3r^{3}}\right) \right)
- \frac{3
\kappa a}{4} \frac{x_{1}x_{i}}{r^{3}}\left( 1-\frac{a^{2}}{r^{2}}\right) .
\end{eqnarray}
This flow field is depicted in \reftext{Fig.~\ref{fig:traction_contours}}(b) for
$\kappa = 1.5, a~= 1$ along with the contours of the (exact) tangential
stress on the particle surface. As expected, the top half of the sphere
has a large shear stress although there are two small regions of reduced
shear stress surrounding the two stagnation points which lie on the
upper hemisphere. The locations of the stagnation points in the neutral
$\{x_{2}=0\}$ plane can again be obtained algebraically as $(\pm
\cos (\phi ),0,\sin (\phi ))$ where
%
\begin{equation}
\sin (\phi ) = \frac{-3}{20}\kappa \pm \frac{1}{20}\sqrt{9\kappa
^{2} + 80}.
\label{eq:kappa}
\end{equation}
For $\kappa <8/3$, both choices of sign in \reftext{\eqref{eq:kappa}} lead to real
arcsines and there is a recirculation region in the flow; in this case
the lower half of the sphere has a region of moderate tangential stress
at the pole and a stagnation point where the incoming flow divides. It
is interesting that for $\kappa $ in this range the fluid immediately
below the body moves in the negative $x_{1}$-direction although the
background flow there was in the positive $x_{1}$-direction. As in the
previous case, this detailed description of the initial flow field
predicts effectively the subsequent evolution, as depicted in
\reftext{Fig.~\ref{fig:movie_with_tracers}} and with outlines given in
\reftext{Fig.~\ref{fig:outlines}}. In particular, we note the persistence of the
recirculation region against the lower portion of the body, separated
by sharp features from the rapidly ablating zone on the upper hemisphere and
from the smaller and more slowly ablating region near the lower pole.

Finally, we consider a background shear flow above a plane wall. The
center of the initially spherical particle is placed at $(0,0,1.5)$, the
background flow is given by $\bm{u}^{\infty }(x) = x_{3}\bm{\hat{x}}
_{1}$, and we impose a no-slip condition on the wall at $\{x_{3}=0\}$.
The exact flow fields and tractions are not so convenient here, but it
is known that the flow between the sphere and the wall is slow and in
the positive $x_{1}$-direction, so there is no recirculation region and
only one stagnation point pair (they appear in pairs because of the
fore-aft symmetry) \cite[Fig.~2]{chaoui2003}. The resulting
evolution, depicted in the third row of
\reftext{Fig.~\ref{fig:movie_with_tracers}} and with outlines in
\reftext{Fig.~\ref{fig:outlines}}(c), accordingly results in a geometry with only
one pair of cusps, in contrast to the greater numbers observed in the
previous cases. In particular, the inclusion of a wall results in slower
flow and a smoother, more spherical body shape there as the particle
erodes toward oblivion.

\section{Conclusion}\label{sec:conclusion}

We used the Lorentz reciprocal theorem to derive a completed traction
boundary integral equation which allows for the incorporation of
background and/or wall-bounded flows. Integrals involving the background
velocity and its stress field were performed on an imaginary bounding
sphere or hemisphere, and the integration was carried out exactly for
the case of linear background flows. We used the method to study two
applications, first to investigate at a greater level of detail the
tractions on glancing and reversing trajectories of ellipsoids near a
wall, and then to study the erosion of bodies by a viscous flow wherein
the surface is ablated at a rate proportional to the local viscous shear
stress. Sharp features were found to develop at stagnation points.
Moreover, the full evolution of the particle geometry can be largely
predicted from an understanding of the flow past the initial shape.
These features of viscous erosion are expected to be generic and
observable for more complicated initial body geometries. An initially
spherical body was found to reduce to a shape reminiscent of Pironneau's
drag-minimizing profile, but we then showed that a self-similar
reduction of an axisymmetric body under constant shear stress occurs
only for a sphere or a cone. 
The surface area was found to vanish in finite time
with a linear scaling $(t-t_{f})$, in contrast to the power law scaling observed at high Reynolds number
\cite{ristroph2012,moore2013self}. 
The presence of a nearby wall was found to have
a smoothing effect, as it reduced the number of stagnation points
appearing on the initial body surface.

The boundary integral equation presented here addresses any background
flow $(\bm{u}^{\infty }, p^{\infty })$ satisfying $ -\nabla p^{\infty
} + \mu \nabla^{2}\bm{u}^{\infty }=\bm{0} $ and $\nabla \cdot \bm{u}
^{\infty} = 0$ throughout the fluid domain. That is, we have considered
the everywhere regular Stokes background flows. The restriction to
regular Stokes flows is not essential and our argument can be extended
to cover, \emph{e.g.}, the problem of finding the tractions on a rigid
body which translates with specified velocity near a point source or a
point force. The extension to several particles is also straightforward.

We found an especially convenient formula in the case of linear
background flows; it would be valuable to carry out these computations
for the case of flows with higher moments as well. A~greater and more
long-term project is to remove the restriction to rigid-body motion.
This is a more profound mathematical challenge because the rigid-body
hypothesis is needed for the essential equation \reftext{\eqref{eq:needs_rbm}}.
In the meantime, problems involving bodies which are changing shape only
slowly relative to a background velocity field might be effectively
handled in a quasi-steady manner, as we have pursued for the problem of
viscous erosion. Other directions of current investigation include
studying the shape dynamics and associated trajectories of force- and
torque-free eroding bodies, and a more detailed analysis of non-smooth
surfaces. The low Reynolds number analogue of the process of melting or
dissolution in flow, independent of shear stress
\cite{rycroft2016asymmetric}, is another appealing direction for future
study.

\section{Acknowledgements}
We gratefully acknowledge helpful conversations with Nick Moore and
Gwynn Elfring. Funding for this research was provided by the
NSF (grant numbers DMS-1056327 and DMR-1121288 (MRSEC)) and the
Wisconsin Alumni Research Foundation.

\appendix

\section{Formulas for fundamental singularities in a half-space}
\label{appA}

Formulas for the fundamental solutions of the Stokes equations in a
half-space have been derived elsewhere by various means: using the
Lorentz reflection procedure \cite{lorentz1896,kk91}, a~Fourier
argument \cite{blake1971,bc74}, brute--force differentiation
\cite{sl12}, and a Papkovich--Neuber potential
\cite{gimbutas2015simple}. In this paper we use formulas obtained
following the Lorentz method. The procedure is as follows. Given a flow
$(\bm{u},p)$ which satisfies the Stokes system on all of $\mathbb{R}
^{3}$ except at a source point $\bm{y}$ above a plane wall at
$\{x_{3} = 0\}$, we define two new solutions $(\bm{v},q)$ and
$(\bm{u}^{*}, p^{*})$ via
%
\begin{gather}
\label{eq:uhatdef}
v_{i} = -\beta_{ij}u_{j} - 2x_{3}\frac{\partial u_{3}}{\partial {x
_{i}}}+x_{3}^{2} \nabla^{2} u_{i},
\ \ \
q = p + 2x_{3}\frac{\partial p}{\partial {x_{3}}} - 4\frac{\partial u
_{3}}{\partial {x_{3}}},
\end{gather}
and
%
\begin{gather}
\label{eq:ustardef}
u^{*}_{i}(\bm{x}) = \beta_{ij} v_{j}(\bm{\beta} \bm{x}),
\ \ \
p^{*}(\bm{x}) = q(\bm{\beta} \bm{x}).
\end{gather}
Here $\beta_{ij} = \delta_{ij} - 2\delta_{3i}\delta_{3j}$ denotes the
operator of reflection over the wall. Both $(\bm{v},q)$ and
$(\bm{u}^{*}, p^{*})$ are solutions of the Stokes equations, and
moreover the latter flow is the required correction to $\bm{u}$ on the
wall: $(\bm{u}+\bm{u}^{*},p+p^{*})$ satisfies the no-slip condition at
$\{x_{3}=0\}$ and has the same singular behavior as $(\bm{u},p)$ at the
source point. Applying these formulas to the point force, point torque,
and stresslet given in \reftext{Table~\ref{tbl:free_singularities}}, and then
simplifying to reduce the algebraic operations required to evaluate the
formulas, we obtain the wall-bounded tensors printed in \reftext{Table~\ref{tbl:wall_singularities}}. We are not aware of any previously
published expression for the stress field of the wall-bounded stresslet.

\section{Useful integration identities}\label{appB}

The integrals on the right-hand side of \reftext{\eqref{eq:general_nowall_shear}} may be evaluated using the identities
%
\begin{gather}
\int_{S} \frac{x_{i}x_{j}}{|\bm{x}|^{4}}dS_{\bm{x}} = \frac{4\pi }{3}
\delta_{ij},
\\
\int_{S} \frac{x_{i}x_{j}x_{k}x_{\ell }}{|\bm{x}|^{6}}dS_{\bm{x}} = \frac{4
\pi }{15}(\delta_{ij}\delta_{k\ell }+\delta_{ik}\delta_{j\ell }+
\delta_{i\ell }\delta_{jk}),
\\
\frac{1}{|\bm{x} - \bm{y}|^{p}} = \frac{1}{|\bm{x}|^{p}} + p\frac{x
_{i}y_{i}}{|\bm{x}|^{p+2}} + \mathcal{O}(1/|\bm{x}|^{p+2}).
\end{gather}
For a linear background flow, $u^{\infty }(\bm{x})=A\bm{x}$, in the
limit where the radius of $S$ increases to infinity, we have
%
\begin{gather}
\frac{\mu }{8\pi }\int_{S} u_{i}^{\infty }(\bm{x}){\hat{n}}_{m}(
\bm{x})T^{STR}_{ijkm}(\bm{x},\bm{y})n_{k}(\bm{y}) dS_{\bm{x}}
= \frac{-3
\mu }{5}(A_{jk}+A_{kj})n_{k}(\bm{y}),
\\
\frac{c\mu }{8\pi }\int_{S} u_{i}^{\infty }(\bm{x}){\hat{n}}_{m}(
\bm{x})C^{STR}_{ijm}(\bm{x},\bm{y})dS_{\bm{x}}
= \frac{3c\mu }{10}(A
_{jk}+A_{kj})z_{k}(\bm{y}) + \frac{c\mu }{2}(A_{jk}-A_{kj})y_{k}
\label{eq:int_c_str},
\\
-\frac{1}{8\pi } \int_{S} f_{i}^{\infty }(\bm{x}) T_{ijk}(\bm{x},
\bm{y})n_{k}(\bm{y}) dS_{\bm{x}}
=\frac{-2\mu }{5}(A_{jk}+A_{kj})n
_{k}(\bm{y}),
\\
- \frac{c}{8\pi }\int_{S} f_{i}^{\infty }(\bm{x})C_{ij}(\bm{x},
\bm{y})dS_{\bm{x}}
= \frac{c\mu }{5}(A_{jk}+A_{kj})z_{k}(\bm{y}).
\label{eq:int_c_vel}
\end{gather}

\def\thetable{B.3}
\begin{table}
\caption{Formulas for three singularity solutions of the Stokes equations in a
fluid with no-slip boundary at $\{x_{3}=0\}$. Here ${\bm{X}} = \bm{x}-\bm{y}$
where $\bm{x}$ is the observation point and $\bm{y}$ is the location of the
singularity above the wall. We write $\beta_{ij} = \delta_{ij} -
2\delta_{3i}\delta_{3j}$ for the reflection operator and $\hat{X}_{i} =
\beta_{ij}x_{j} - y_{i}$. In the denominators we have written $R = |\bm{X}|$ and
$\hat{R} = | \bm{\hat{X}}|$. It is interesting that the velocity tensor of the
stresslet does not match the stress tensor of the stokeslet as in free space;
indeed the stress tensor of the wall-bounded stokeslet does not vanish on the
wall.}
\label{tbl:wall_singularities}
{\small
\begin{tabular*}{\textwidth}{ll}
\hline
Stokeslet velocity $G^{\text{half}}_{ij}$ & $\displaystyle \frac{\delta _{ij}}{R}-\frac{\delta _{ij}}{\hat{R}} + \frac{X_{i} X_{j}}{R^{3}}- \frac{X_{i} X_{j}}{\hat{R}^{3}} -\frac{2\delta _{ij} y_{3}x_{3}}{\hat{R}^{3}} + \frac{6 y_{3}x_{3}\hat{X}_{j}\beta _{i\ell }\hat{X}_{\ell }}{\hat{R}^{5}}$
\\ [2mm]
Stokeslet pressure $G^{\text{half, }P}_{j}$& $\displaystyle 2\frac{X_{j}}{R^{3}} - 2\frac{X_{j}}{\hat{R}^{3}} -12Y_{3}\frac{\hat{X}_{3}\hat{X}_{j}}{\hat{R}^{5}}$
\\ [2mm]
Stokeslet stress $G^{\text{half, }STR}_{ijk}$&
$\displaystyle
-6\frac{X_{i}X_{j}X_{m}}{R^{5}}
+ \frac{6}{\hat{R}^{5}}\left[
X_{i}X_{j}X_{m}-2\delta _{im}y_{3}^{2}\hat{X}_{j} + \delta _{3m}y_{3}(X_{i}X_{j} + \beta _{it}\hat{X}_{t}\hat{X}_{j}) + \delta _{3i}y_{3}(X_{j}X_{m} + \hat{X}_{j}\beta _{mt}\hat{X}_{t})\right.$
\\ [2mm]
& \; $\displaystyle
{}+ \left.x_{3}y_{3}(\delta _{ij}\beta _{mt} + \delta _{jm}\beta _{it} + \beta _{jm}\beta _{it} + \beta _{ij}\beta _{mt})\hat{X}_{t}
\right] - 30\frac{x_{3}y_{3}\hat{X}_{j}(\beta _{i\ell }\beta _{mt} + \beta _{m\ell }\beta _{it})\hat{X}_{\ell} \hat{X}_{t}}{\hat{R}^{7}}$
\\   \noalign{\vspace{9pt}}

Rotlet velocity $R^{\text{half}}_{im}$ & $\displaystyle \frac{\epsilon _{imk} X_{k}}{R^{3}} - \frac{\epsilon _{imk} X_{k}}{\hat{R}^{3}} -\frac{6\epsilon _{3mk}\hat{X}_{k}x_{3}\beta _{i\ell }\hat{X}_{\ell} }{\hat{R}^{5}}$
\\ [2mm]
Rotlet pressure $R^{\text{half, }P}_{m}$& $\displaystyle 12\frac{\epsilon _{3mk}\hat{X}_{k}\hat{X}_{3}}{\hat{R}^{5}} $
\\ [2mm]
Rotlet stress $R^{\text{half, }STR}_{imp}$& $\displaystyle
-3\frac{\epsilon _{imk}X_{k}X_{p} + \epsilon _{pmk}X_{k}X_{i}}{R^{5}} +\frac{60}{\hat{R}^{7}} x_{3}\beta _{i\ell }\hat{X}_{\ell} \epsilon _{3mk}\hat{X}_{k} \beta _{ps}\hat{X}_{s}
+\frac{3}{\hat{R}^{5}} \left(
\epsilon _{imk}\hat{X}_{k}\beta _{ps}\hat{X}_{s} + \epsilon _{pmk}\hat{X}_{k}\beta _{is}\hat{X}_{s}
+ 2x_{3}\epsilon _{im3}\beta _{ps}\hat{X}_{s} \right.
$
\\ [2mm]
& \; $\displaystyle {}
+ \left.2x_{3}\epsilon _{pm3}\beta _{is}\hat{X}_{s} -2\delta _{3p}\beta _{i\ell }\hat{X}_{\ell} \epsilon _{3mk} \hat{X}_{k}-2\delta _{3i}\beta _{p\ell }\hat{X}_{\ell} \epsilon _{3mk} \hat{X}_{k}
+4y_{3}\delta _{ip}\epsilon _{3mk}\hat{X}_{k}
-2x_{3}\beta _{i\ell }\hat{X}_{\ell} \epsilon _{3mp} -2x_{3}\beta _{p\ell }\hat{X}_{\ell} \epsilon _{3mi} \right)
$
\\  \noalign{\vspace{9pt}}

Stresslet velocity $T^{\text{half}}_{ijk}$ & $\displaystyle -6\frac{X_{i}X_{j}X_{k}}{R^{5}} +6\frac{X_{i}\hat{X}_{j}\hat{X}_{k}}{\hat{R}^{5}} - \frac{12x_{3}\left( x_{3}\delta _{jk}\beta _{im}\hat{X}_{m} -y_{3}\beta _{ik}\hat{X}_{j}-y_{3}\beta _{ij}\hat{X}_{k}\right) }{\hat{R}^{5}} - 60x_{3}y_{3}\frac{\hat{X}_{j}\hat{X}_{k}\beta _{im}\hat{X}_{m}}{\hat{R}^{7}}$
\\ [2mm]
Stresslet pressure $T^{\text{half, }P}_{jk}$ & $\displaystyle 4\delta _{jk}\left( \frac{1}{R^{3}}+\frac{1}{\hat{R}^{3}}\right)
-12\frac{X_{j}X_{k}}{R^{5}}+12\frac{\hat{X}_{j} \hat{X}_{k}}{\hat{R}^{5}} -24y_{3}\frac{\delta _{3k}\hat{X}_{j} + \delta _{3j}\hat{X}_{k}}{\hat{R}^{5}} +24\delta _{jk}\frac{x_{3}\hat{X}_{3}}{\hat{R}^{5}} + 120y_{3}\frac{\hat{X}_{3}\hat{X}_{j}\hat{X}_{k}}{\hat{R}^{7}}$
\\ [2mm]
Stresslet stress $T^{\text{half, }STR}_{ijkm}$&
$\displaystyle
-4\delta _{im}\delta _{jk}\left( \frac{1}{R^{3}}+\frac{1}{\hat{R}^{3}}\right)
-\frac{6}{R^{5}}\left( \delta _{ij}X_{k}X_{m} + \delta _{ik}X_{j}X_{m} + \delta _{jm}X_{i}X_{k} + \delta _{km}X_{i}X_{j}\right)
+\frac{60}{R^{7}}X_{i}X_{j}X_{k}X_{m}$
\\ [2mm]
& \;$\displaystyle{}
+\frac{6}{\hat{R}^{5}}\left[ 4\delta _{im}y_{3}(\delta _{3k}\hat{X}_{j} + \delta _{3j}\hat{X}_{k})+4x_{3}y_{3}(\delta _{im}\delta _{jk}+\beta _{ik}\beta _{jm} + \beta _{ij}\beta _{km})
+ \beta _{it}\hat{X}_{t}(\beta _{jm}\hat{X}_{k} + \beta _{km}\hat{X}_{j} - 4\delta _{3m}x_{3}\delta _{jk})
\right. $
\\ [2mm]
& \; $\displaystyle {}
+ \left.\beta _{mt}\hat{X}_{t}(\beta _{ij}\hat{X}_{k} + \beta _{ik}\hat{X}_{j} - 4\delta _{3i}x_{3}\delta _{jk})
\right]
+\frac{60}{\hat{R}^{7}} \left[ 2\delta _{im}y_{3}^{2}\hat{X}_{j}\hat{X}_{k} + 2\delta _{jk}x_{3}^{2}\beta _{it}\hat{X}_{t}\beta _{mp}\hat{X}_{p}
- \hat{X}_{j}\hat{X}_{k}\beta _{it}\hat{X}_{t}\beta _{mp}\hat{X}_{p}
\right. $
\\ [2mm]
&\;$\displaystyle {}
-\left.2x_{3}y_{3}\beta _{mp}\hat{X}_{p}(\beta _{ik}\hat{X}_{j} + \beta _{ij}\hat{X}_{k}) - 2x_{3}y_{3}\beta _{it}\hat{X}_{t}(\beta _{km}\hat{X}_{j} + \beta _{jm}\hat{X}_{k})\right]
+\frac{840}{\hat{R}^{9}}x_{3}y_{3}\hat{X}_{j}\hat{X}_{k}\beta _{it}\hat{X}_{t}\beta _{mp}\hat{X}_{p}
$\\
\hline
\end{tabular*}
}
\end{table}

\end{document}